\documentclass[fleqn,usenatbib]{mnras}

\usepackage{newtxtext,newtxmath}
\usepackage{lipsum}  
\usepackage{subcaption}

\usepackage[T1]{fontenc}
\usepackage{hyperref}
\DeclareRobustCommand{\VAN}[3]{#2}
\let\VANthebibliography\thebibliography
\def\thebibliography{\DeclareRobustCommand{\VAN}[3]{##3}\VANthebibliography}

\usepackage{graphicx}	
\usepackage{amsmath}	
\usepackage{xcolor}

\title[Evaluating bulk flow estimators]{Evaluating bulk flow estimators for CosmicFlows--4 measurements}

\author[A M. Whitford et al.]{
Abbé M. Whitford,$^{1}$\thanks{E-mail:abbe.whitford@gmail.com}
Cullan Howlett,$^{1}$
Tamara M. Davis$^{1}$
\\
$^{1}$School of Mathematics and Physics, The University of Queensland, Brisbane, QLD 4072, Australia\\
}

\date{Accepted XXX. Received YYY; in original form ZZZ}

\pubyear{2015}

\begin{document}
\label{firstpage}
\pagerange{\pageref{firstpage}--\pageref{lastpage}}
\maketitle

\begin{abstract}
For over a decade there have been contradictory claims in the literature about whether the local bulk flow motion of galaxies is consistent or in tension with the $\Lambda$CDM model. While it has become evident that systematics affect bulk flow measurements, systematics in the estimators have not been widely investigated. In this work, we thoroughly evaluate the performance of four estimator variants, including the Kaiser maximum likelihood estimator (MLE) and the minimum variance estimator (MVE). We find that these estimators are unbiased, however their precision may be strongly correlated with the survey geometry. Small biases in the estimators can be present leading to underestimated bulk flows, which we suspect are due to the presence of non-linear peculiar velocities. The uncertainty assigned to the bulk flows from these estimators is typically underestimated, which leads to an overestimate of the tension with $\Lambda$CDM. We estimate the bulk flow for the CosmicFlows--4 data and use mocks to ensure the uncertainties are appropriately accounted for. Using the MLE we find a bulk flow amplitude of $408\pm165 \mathrm{km s}^{-1}$ at a depth of $49\, \mathrm{Mpc} h^{-1}$, in reasonable agreement with $\Lambda$CDM. However using the MVE which can probe greater effective depths, we find an amplitude of $428\pm108 \mathrm{km s}^{-1}$ at a depth of $173\, \mathrm{Mpc} h^{-1}$, in tension with the model, having only a 0.11\% probability of obtaining a larger $\chi^2$. These measurements appear directed towards the Great Attractor region where more data may be needed to resolve tensions.
\end{abstract}




\section{Introduction}

Local fluctuations in the underlying matter density of the Universe source the gravitational motions of galaxies. These motions create local velocity flows that drag galaxies towards each other. A measurement of the average of these motions is called the \emph{bulk flow}. The bulk flow in a particular volume gives us a picture of the direction and amplitude of the overall flow of matter in that region. Because the bulk flow arises due to the Large Scale Structure (LSS), it is not only a useful tool that can be used to map motions in the local Universe but it also allows us to test models of cosmology. The bulk flow is related to $\Omega_m$, the total matter energy density due to baryons and dark matter, and $\sigma_8$, the variance in matter fluctuations in spheres of radius $8$ Mpc $h^{-1}$. We can also use bulk flow measurements to test theories of gravity, such as General Relativity.

The bulk flow is particularly interesting at the present time due to tensions in bulk flow measurements over the last decade or more. A number of measurements have claimed the measured bulk flows are in some tension with the current concordance model of cosmology, $\Lambda\text{CDM}$ \citep{kashlinsky2008measurement, watkins2009consistently, feldman2010cosmic, watkins2015large, peery2018easily, howlett2022sloan, watkins2023analyzing}. In contrast, a number of measurements have found they are in agreement \citep{nusser2011cosmological, hong20142mtf, ma2014estimation, hoffman2015cosmic, scrimgeour20166df, qin2018bulk, qin2021cosmic}. Interestingly, many of the measurements that are in tension with the $\Lambda$CDM model measure the bulk flow on larger scales than those that appear in agreement.
Given these disagreements, the aim in this paper is to make the most robust measurement yet of the bulk flow using the latest CosmicFlows--4 catalogue, while also considering the precision and accuracy of the bulk flow estimators we employ in order to properly determine the consistency of the measurement with the $\Lambda\text{CDM}$ model. In particular, it is interesting to test if additional peculiar velocity data resolves or worsens tensions in bulk flow measurements. 

Formally the bulk flow $\mathbf{B}$ is a measurement of the average peculiar velocity of galaxies in a given survey volume $V$, 
\begin{equation} \label{eq::definition_bulkfflow}
    \mathbf{B} = \frac{1}{V} \int_V \mathbf{v}(\mathbf{r}) d^3r,
\end{equation}
where $\mathbf{v}(\mathbf{r})$ represents the 3--dimensional peculiar velocity (PV) field. The PV of a galaxy refers to its motion due to interactions with local gravitational fields, rather than motion due to the expansion of space. At present, we are only able to measure the radial component of the PV of a galaxy. Therefore the components of $\mathbf{B}$ along each coordinate-axis in a volume, $B_{i}$, may be estimated as a weighted average of radial PV measurements,
\begin{equation}
    B_{i} \approx \frac{1}{V} \int u(\mathbf{r}) \hat{n}_{i} w_{i}(\mathbf{r}) d^3r \approx \sum_m u_m \hat{n}_{i, m} w_{i,m} = \tilde{B}_{i}.
\end{equation}
In the above equation, $\tilde{B}_{i}$ is the estimator for $B_{i}$, where $\mathbf{B} = \sum_{i} B_{i}  \hat{\mathbf{x}}_{i}$. The radial PV field we measure is given by $u(r) = \mathbf{v}(\mathbf{r}) \cdot \hat{\mathbf{r}}$ where $\hat{r}$ is the unit vector for the line-of-sight. Above, $u_m$ represents the observed PV for the \emph{m}th galaxy in the sample, $w_{i,m}$ represents the weight applied to the \emph{m}th galaxy for the $i$--coordinate direction (given some estimator used to determine the optimal weighting scheme) and $\hat{n}_{i,m}$ gives the projection of the radial PV onto the $i$ coordinate direction, $\hat{n}_{i,m} = \hat{\mathbf{r}}_{m} \cdot \hat{\mathbf{x}}_{i}$. 

Alternatively, one may attempt to reconstruct the full 3--dimensional PV field, using a method such as the Wiener Filter \citep{zaroubi1994wiener}, then sum the 3--dimensional PVs to obtain a bulk flow estimate. An issue that may arise with this method is the need for prior information about the underlying cosmology of the Universe to do the reconstruction. This may dominate the underlying bulk flow signal if the data used to do the reconstruction is noisy. In general however, different methods of measuring the bulk flow may obtain results that are not directly comparable because they define the true bulk flow of the data differently \citep{nusser2016methods}. Assumptions in bulk flow estimators needs to be taken carefully into consideration, along with potential systematics, before comparing a bulk flow estimate to theory. 

As stated previously we aim to make a measurement of the bulk flow, primarily using the CosmicFlows--4 catalogue, the largest combined dataset of PVs to date \citep{tully2023cosmicflows}. This dataset consists of combined data from the Sloan Digital Sky Survey peculiar velocity catalogue (SDSS; \citealt{howlett2022sloan}), the CosmicFlows IV Tully--Fisher catalogue (CF4TF; \citealt{kourkchi2020cosmicflows}) and the 6-Degree Field Galaxy peculiar velocity survey (6dFGSv; \citealt{springob20146df}). The dataset also contains PVs measured from other low-redshift objects that can be used to construct a distance ladder and together make a total of 55,877 measurements. This combined sample will also probe more deeply than other combined samples; this is mainly due to the new SDSS sample which contains peculiar velocities for galaxies as deep as $z \sim 0.1$. However, the peculiar `top-heavy' geometry also offers a potential route for systematics to enter. A discussion for the datasets and mocks that we test our methodologies on can be found in section~\ref{sec::data_and_realisticmocks}. We also present results for measurements from the individual SDSS, 6dFGSv and CF4TF datasets, as well as their combination when other low redshift objects are not included.

Prior to this in section~\ref{sec::systematics_in_bulkflow_estimators} we discuss systematics in estimators for the bulk flow and the results of our own tests of the performance of popular bulk flow estimator methodologies. We apply these estimators to a range of mocks (ranging from simplistic to sophisticated in nature) for the datasets of interest in order to evaluate their performance thoroughly in section~\ref{sec::mock_performance_eval}. Our aim here is to closely investigate systematics in the Kaiser Maximum Likelihood Estimator \citep[Kaiser MLE;][]{kaiser1988theoretical} and the Minimum Variance Estimator by \cite{watkins2009consistently} (Watkins MVE). We also investigate the Maximum Likelihood approach suggested in \cite{nusser2014inconsistency} (hereon Nusser MLE) and the modifications to the Minimum Variance Estimator in \cite{peery2018easily} (hereon Peery MVE). We discuss the pros and cons of these methods, and which methods we expect are best when applied to survey data. This is an important sanity-check for solving tensions in bulk flow measurements, especially in light of the recent results by \cite{watkins2023analyzing} who claim to make a bulk flow measurement also using CosmicFlows--4, that is in significant tension with the $\Lambda$CDM model. Our tests on mock data show that the application of the MVE and the Kaiser MLE to the CosmicFlows--4 data tend to obtain a slightly underestimated measurement of the bulk flow, however our results also show that the uncertainty on the measurement is also likely underestimated, which overestimates the amount of tension between $\Lambda$CDM with the measurement. Furthermore, we explore the effects of the zero--point calibrations applied to the individual datasets that compose CosmicFlows--4 to show how this changes the measured bulk flow amplitude.

Our results from applying the bulk flow measurement techniques to realistic mocks and the various datasets are shown in section~\ref{sec::results}, along with a comparison between the results from the data and mocks with theoretical expectations to quantify the level of tension with $\Lambda$CDM. Finally in section~\ref{sec::conclusions} we conclude this paper with a discussion of recommendations for future work regarding bulk flow measurements and potential ways to improve existing methods of measuring the bulk flow to resolve tensions between datasets and robustly test the $\Lambda$CDM model.

\section{Datasets and mock catalogues}\label{sec::data_and_realisticmocks}

Properties of the CosmicFlows--4 (CF4) data and the largest subsets composing it (SDSS, 6dFGSv and CF4TF) are discussed in this section and summarised in Table~\ref{tab:surveydata_properties}, Figure~\ref{fig:data_selectionfunctions} and Figure~\ref{fig:data_skycoverage}. The mocks we use for the CF4 data, by combining mocks for SDSS, 6dFGSv and the CF4TF datasets and which capture the majority of the data in CF4 and the survey geometry, are also discussed in this section. 

For all the CF4 mocks and data, peculiar velocities are estimated from log-distance ratios using the \cite{watkins2015unbiased} estimators. Furthermore, all galaxy distances calculated for determining weights in any of the estimators or the effective depth of the data use the observed redshifts of galaxies. 

\subsection{The Sloan Digital Sky survey peculiar velocity catalogue}

\subsubsection{The data}

The Sloan Digital Sky Survey (SDSS) peculiar velocity catalogue \citep{howlett2022sloan} is a set of 34,059 peculiar velocities on the Fundamental Plane \citep[FP, ][]{djorgovski1987fundamental}, comprising the largest individual set of peculiar velocities to date. This dataset probes more deeply than the other samples we combine with to a redshift of $z=0.1$ but only covers only a small region of the sky in the SDSS northern galactic cap. The mean uncertainty on each FP distance measurement is $\sim23\%$. This dataset contains group redshifts for galaxies from the same underlying dark matter haloes that allow for the galaxy PVs to be measured more accurately. Peculiar velocities are obtained by building on \cite{said2020joint} and fitting the Fundamental Plane parameters extracted from the SDSS Data Release 14 \citep{abolfathi2018fourteenth}.

\subsubsection{The mocks}

\cite{howlett2022sloan} also provide a set of 2048 realistic mocks for this dataset. The mocks reproduce the sky mask and selection function of the SDSS PV data and were designed to capture all aspects of the real data, including selection effects, measurement errors and the effects of cosmic variance. More details can be seen in \cite{howlett2022sloan}, and the mocks can be found at \href{https://zenodo.org/record/6640513}{https://zenodo.org/record/6640513}. 

\subsection{The CosmicFlows IV Tully--Fisher peculiar velocity catalogue}

\subsubsection{The data}

The CosmicFlows IV Tully--Fisher catalogue \citep[CF4TF, ][]{kourkchi2020cosmicflows} consists of 9534 peculiar velocities drawn from the Tully--Fisher plane relation \citep[TF, ][]{tully1977new} that are mostly within a redshift of $z \sim 0.05$ and cover the sky entirely (excluding the galactic plane). Unlike SDSS, this dataset is shallower, but it is apparent in Figure~\ref{fig:data_selectionfunctions} that the selection function of this dataset is complimentary to the SDSS dataset by having a high density of objects at low redshift. For this dataset, the H I linewidths and fluxes used for the TF relation parameters come from measurements by the All Digital H I catalogue \citep[ADHI, 78\%,][]{courtois2009extragalactic}, the Aricebo Fast ALFA Survey \citep{haynes2011arecibo, haynes2018arecibo}, the Springbob/Cornell H I catalogue \citep{springob2005digital} and the Pre Digital H I catalogue \citep{fisher1981upper, huchtmeier1989hi}. Photometry data for the galaxies is taken from the SDSS DR12 data release \citep{york2000sloan}. More information can be found within \cite{kourkchi2020cosmicflows}.

\subsubsection{The mocks}

For this dataset mocks have been produced by \cite{qin2021cosmic}, which reproduce the survey geometry and selection function for the CF4TF data. The mock sampling algorithm used here first by \cite{qin2021cosmic} to generate the mocks is the same as that used to produce the SDSS mocks by \cite{howlett2022sloan} with the corresponding survey properties. As such, the mocks for the CF4TF data are also designed to capture all the aspects of the real data including cosmic variance, measurement errors and selection effects.

\subsection{The 6-degree Field Galaxy survey PV catalogue}

\subsubsection{The data}

This 6dFGSv dataset \citep{springob20146df} consists of another 8885 galaxies with peculiar velocities measured from the FP relation. This sample covers the entire Southern sky (excluding the plane of the Milky Way) and is thus complimentary to the SDSS and CF4TF data. This dataset is deeper than CF4TF with more redshifts at $z \sim 0.03 - 0.056$ but is shallower than the SDSS sample and has greater sky coverage. The typical uncertainties on each galaxy distance from this FP are around $\sim26\%$.
The spectroscopic observations of the FP galaxies in this sample were made with the UK Schmidt Telescope as part of the 6dF survey \citep{Jones20096df}, and photometric observations from the Two Micron All-Sky Survey (2MASS) Extended Source Catalog \citep{jarrett20002mass}. The data used here (and within the full CF4 release) is a reprocessing of the original data including a modified correction for the selection function when extracting distances from the Fundamental Plane. The reprocessing affects only the extraction of distances/velocities, not the publicly available photometric or spectroscopic measurements, and is fully described in \citet{qin2018bulk} and \citet{tully2023cosmicflows}.

\subsubsection{The mocks}

Mocks for the 6dFGSv survey are created following \cite{qin2019redshift}, although with slight modifications as presented in \cite{tully2023cosmicflows} and to align their construction with the SDSS and CF4TF methodology used to produce the mock catalogues in \cite{howlett2022sloan} and \cite{qin2021cosmic}. Again, these reproduce the selection function of the survey and are designed to capture the effects of cosmic variance, measurement errors and selection effects as closely as possible to the real survey.

\subsection{The combined CF4 dataset and mocks}

\subsubsection{The data}

\begin{figure}
    \centering
    \includegraphics[width=0.47\textwidth]{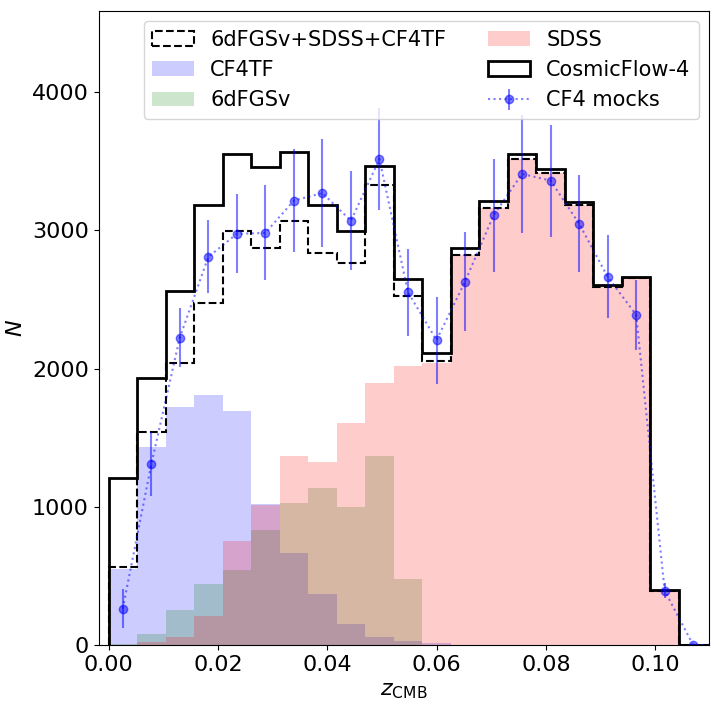}
    \caption{Selection functions of individual data for CF4TF, SDSS and 6dFGSv, and the combination of all. The selection function for CF4 data (including low redshift SNe and TRGB stars and others) is also shown. $z_{\mathrm{CMB}}$ represents the redshift in the CMB frame. Each of the bins has a width of $\Delta z = 0.005$. }
    \label{fig:data_selectionfunctions}
\end{figure}

Figure~\ref{fig:data_selectionfunctions} shows the redshift selection function for each dataset discussed in the previous section, the redshift selection function for the combined data from these datasets that is included in CF4 \cite[this is not all of the 8885 galaxies for 6dFGSv because some of these are removed in the catalogue due to their classification as spiral interlopers in][]{tully2016cosmicflows} and the redshift selection function for the entire CosmicFlows--4 catalogue which includes low-redshift objects used to calibrate the zero--point for the data in \cite{tully2023cosmicflows}. 

\begin{figure*}
    \centering
    \includegraphics[scale=0.5]{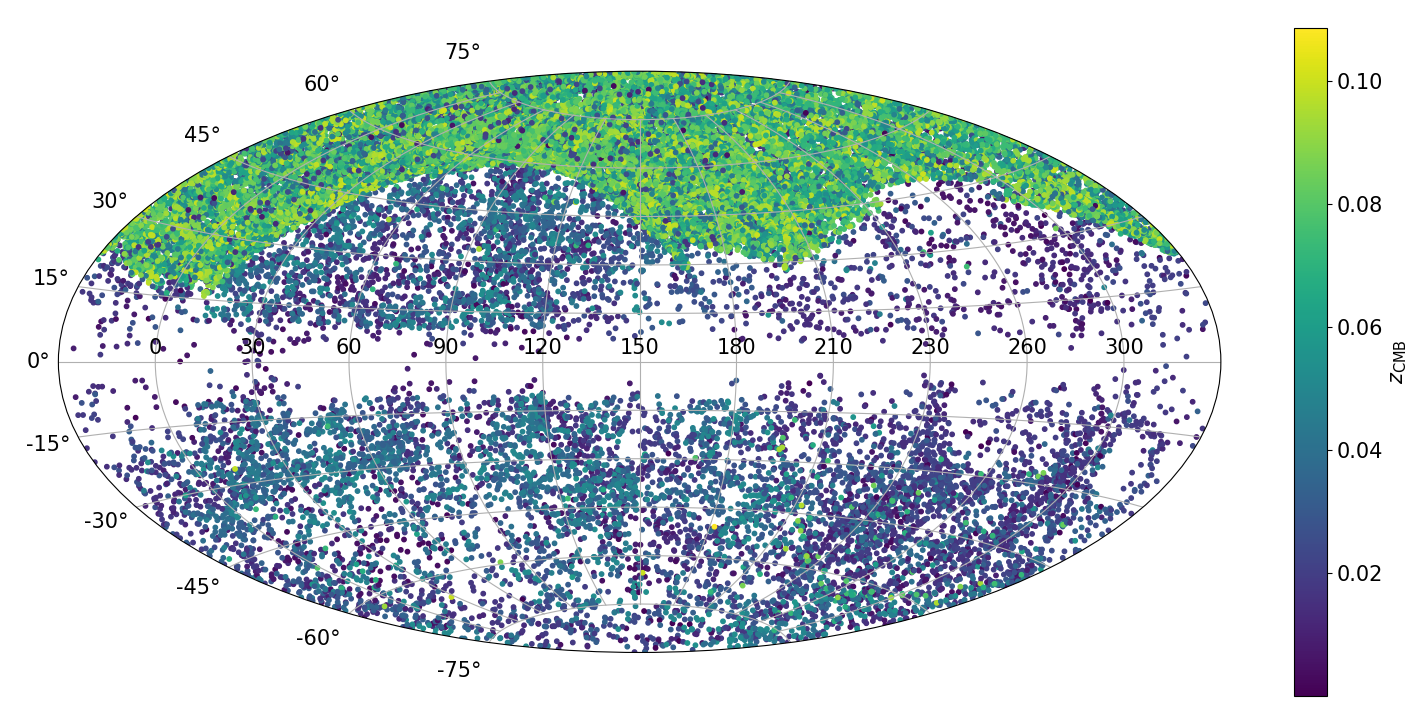}
    \caption{The sky coverage for the entire CosmicFlows--4 dataset shown in galactic coordinates. Each point here represents a group out of the 38008 groups of objects made from the 55877 objects. The color of each point shows the redshift relative to the Cosmic Microwave Background (CMB) reference frame according to the colorbar. }
    \label{fig:data_skycoverage}
\end{figure*}

Figure~\ref{fig:data_skycoverage} shows the sky coverage for the entire CosmicFlows--4 dataset. The distance measurements in CF4 that are not included in SDSS, CF4TF and 6dFGSv include those from Cepheid variable stars \citep{leavitt1912periods}, type Ia SNe \citep{phillips1993absolute} , type II SNe \citep{hamuy2002type}, surface brightness fluctuations (SBFs) in elliptical galaxies \citep{tonry1988new} and tip of the red-giant branch \citep[TRGBs;][]{lee1993tip}.  Further FP and TF measurements are also used from other samples detailed in \citet{tully2023cosmicflows}. These other samples have overlapping galaxies with the SDSS, CF4TF and 6dFGSv datasets already discussed and independent distance measurements for the same galaxy have been averaged in the combined dataset for all CF4. 

Parallax distance measurements and geometric maser distances \citep{humphreys2013toward} help set the absolute distance scale for TRGBs and Cepheid-variable stars. These are used to help provide an absolute distance scale and allow for the zero--point of the datasets to be calibrated using a Bayesian methodology. The zero--point offset of the galaxy peculiar velocities represents the monopole of the field and is degenerate with $H_0$. The need for a correction arises during the FP (Fundamental Plane) and TF (Tully--Fisher) fitting procedures, as it is assumed when measuring these relations that the zero--point of the galaxy PVs is null, which causes an unknown constant offset in the true PVs of the galaxies relative to the measurements that requires correction. In \cite{tully2023cosmicflows}, the overlapping objects of various datasets for FP, TF, SNe Ia, SNe II, and TRGB stars are used to calibrate datasets for galaxy distances measured using the same methodology to each other. Then all of the datasets are tied to an absolute scale set by the SNe Ia objects using overlapping galaxy groups in the samples. More detail can be found in \cite{tully2023cosmicflows}. The process for calibrating the various zero-points carries potential for systematic error, which is one of the things we test in this work.

\begin{table}
	\centering
	\caption{Properties of each peculiar velocity dataset we explore in this work, where $z$ refers to the CMB-frame redshift and the mean error column refers to the mean error on each measured galaxy distance.}
	\label{tab:surveydata_properties}
	\begin{tabular}{ccccc} 
		\hline
		Survey & No. galaxies & $z$ range & Median $z$ & mean error \\
		\hline
		SDSS PVs & 34059 & 0.0033--0.1 & 0.071 & 23\% \\
		CF4TF PVs & 9534 & 0.0--0.064 & 0.018 & 22\% \\
		6dFGS PVs & 8885 & 0.0--0.056 & 0.039 & 26\% \\
            CF4 & 55877 & 0.0-0.1 & 0.051 & 21.5\% \\
		\hline
	\end{tabular}
\end{table}

\subsubsection{The mocks}

We take the mocks from \cite{qin2019redshift, qin2021cosmic, howlett2022sloan} discussed in the previous sections for SDSS, 6dFGSv and CF4TF. These mocks have been all made using the L-PICOLA $n$-body code \citep{howlett2015picola} with the same settings and initial conditions for each mock, such that all the galaxies are assigned to haloes from the same dark matter distribution. They have also been placed in the simulation in such a way that the positions of galaxies relative to the observer mimics the relative positions of the true galaxy surveys; this allows the survey mocks to be stacked together in order to make combined mocks for the datasets and also largely capture the geometry of CF4. We present the performance of the bulk flow estimators applied to the mocks in section~\ref{sec::results}. The blue points in Figure~\ref{fig:data_selectionfunctions} shows the averaged redshift selection function of the CF4 mocks we use, and the error bars show the standard deviation of the number counts in each bin for the mocks.

\section{Systematics in bulk flow measurements}\label{sec::systematics_in_bulkflow_estimators}

\subsection{Overview}

Various difficulties exist in regards to producing accurate PV measurements and subsequently a bulk flow measurement. Firstly, measurements of PVs derived from distances obtained with empirical scaling relations, such as the FP or the TF relation, are noisy and have an uncertainty on the measured galaxy distances that are typically $\sim$ 20\% \citep{strauss1995density}. Type Ia SNe are generally able to obtain more accurate distance measurements with an uncertainty of 5--10\% \citep{fakhouri2015improving,rest2014cosmological}. Large uncertainties on the distance measurements that propagate through to PV calculations lead to a noisier measurement of the bulk flow. However, this measurement noise can be combated by increasing the number of PV measurements. PV measurements may also have a non--Gaussian PDF, although this may be overcome in various ways \citep{watkins2015unbiased, qin2018bulk, qin2021gaussianization, hoffman2021cosmicflows}. In general, the galaxy peculiar velocities may have uncertainties that follow a log-normal distribution. Fortunately, the \cite{watkins2015unbiased} estimator allows for PVs with Gaussian error bars to be estimated from Gaussian distributed log--distance ratios,\footnote{The log--distance ratio $\eta$ of a galaxy is defined as $\eta = \log_{10}{\left(\frac{D(z_{\text{CMB}})}{D(z_{\text{rec}})}\right)}$, where $D(z_{\text{CMB}})$ is the comoving distance to the galaxy computed from a cosmological model given the observed redshift of the galaxy, and $D(z_{\text{rec}})$ is the true comoving distance to the galaxy. A measurement of $\eta$ is typically what is obtained from the Fundamental Plane or Tully--Fisher relation in order to calculate a PV. } under the assumption that the PV $u_m$ obeys $ u_m \ll cz_{m, \mathrm{CMB}}$, where $z_{m, \mathrm{CMB}}$ is the observed redshift of the galaxy from spectroscopy, i.e. the peculiar velocity is much smaller than the recession velocity. Alternatively, the Box Cox transformation also allows for any distribution to be Gaussianised \citep{qin2021gaussianization}. 

A second issue which arises is that only the radial (line--of--sight) component of the galaxy's peculiar motion, $u_m = \mathbf{v} \cdot \hat{\mathbf{r}}_m$ can be measured. This is because the PVs are derived from the galaxy's redshift, which for non--relativistic velocities is only due to the radial component of the galaxy's motion away from the observer. Therefore, the weighting scheme applied to the PVs in order to estimate the bulk flow must be derived in such a way that it as closely as possible captures the 3--dimensional bulk flow from what is effectively 1--dimensional PV data for each galaxy. \cite{nusser2014inconsistency} shows that under the assumption that the 3--dimensional PV field has no curl component, it is possible to entirely gain the 3--dimensional information about the bulk flow from only the radial projection of the PV field, for a full-sky dataset. Results from our tests on mocks show it also is generally possible (on average) when the bulk flow vector can be described as a constant valued vector across the survey volume. Alternatively, one can consider scheme's such as the Wiener Filter \citep{zaroubi1994wiener} to reconstruct the 3--dimensional PV field from the data and thus estimate the bulk flow. However as mentioned previously, for small or noisy samples of PVs the prior information assumed from a cosmological model when employing this method can dominate the underlying signal. This makes the process of comparing a bulk flow thus derived to a cosmological model somewhat circular. 

\begin{figure}
    \centering
    \includegraphics[width=0.48\textwidth]{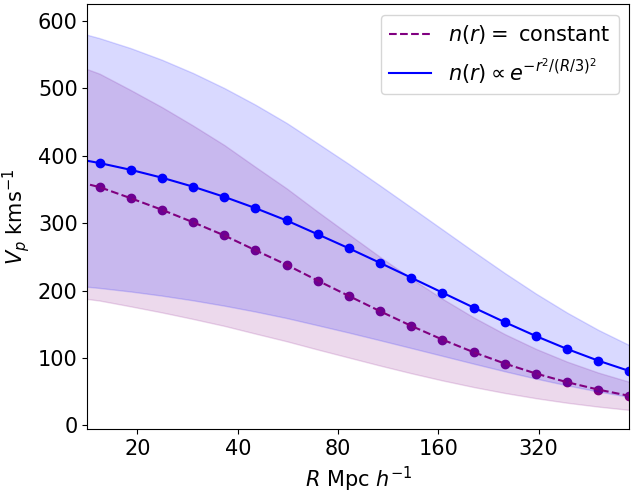}
    \caption{The theoretical bulk flow $V_p$ as a function of the survey radius $R$ for two survey geometries. The shaded regions show the $1-\sigma$ uncertainty on the theoretical bulk flow due to cosmic variance. For each line plotted, the legend shows the functional form of the number density of objects $n(r)$ for the survey selection function. The predictions here have been calculated using the equations and methods given in \protect\cite{andersen2016cosmology}.  The predicted bulk flow has been computed using a grid of $500^3$ points, with $H_0 = 67.74\,\mathrm{km s}^{-1} \mathrm{Mpc}^{-1}$, $\Omega_m = 0.3089$ and with $\Omega_{\Lambda} = 1.0 - \Omega_m$ for the standard $\Lambda\text{CDM}$ model. } 
    \label{fig:theoreticalbulk_vary_surveygeometry}
\end{figure}

\cite{andersen2016cosmology} looked into how the survey geometry, in particular the sky mask applied to the survey data, can affect bulk flow measurements. Using simulations, their paper shows that the predicted theoretical bulk flow needs to take into account the geometry of the PV galaxy survey data before making a comparison of a bulk flow measurement to theoretical predictions. They also propose how the bulk flow from theoretical predictions can be more accurately computed to take into account the survey geometry. We demonstrate the effect of the survey geometry on the theoretical bulk flow prediction using their methods in Figure \ref{fig:theoreticalbulk_vary_surveygeometry}. This demonstrates how the theoretical bulk flow should change for different survey selection functions as a function of survey depth. Furthermore \cite{andersen2016cosmology} show with simulations that under sampling of the PV field (i.e., a number of measurements $ < 500$) affects the bulk flow amplitude and gives resulting bulk flow measurements that do not necessarily agree with expectation from theory as the variance of the measured bulk flow increases; in the case of small sample sizes, a measured bulk flow needs to be compared to a prediction from simulations or mock catalogues. Overall, it is most important to take care if directly comparing the bulk flow amplitude and cosmic variance uncertainty to a measurement, especially when placing a coordinate for the amplitude of the measurement on a theory plot similar to that shown in Figure~\ref{fig:theoreticalbulk_vary_surveygeometry}. 

Finally, the last issue we will discuss here and which is a main focus of investigation in this work is related to the underlying performance of the estimators used to measure the bulk flow. Various popular estimators exist in the literature to derive bulk flow measurements, but incorporate assumptions about the nature of the measured bulk flow, or the data, that can have an affect on their ability to 1) return a realistic precision on the bulk flow from the data and and 2) return an accurate bulk flow measurement compared to the moment of the data we expect the estimator to recover. 

Here in particular we are interested in investigating the precision of the \cite{kaiser1988theoretical} maximum likelihood estimator (Kaiser MLE) which has widely been used through out the literature to measure the bulk flow. Various authors have found the MLE approach does not provide a realistic estimate of the precision on the measured bulk flow moment (the error bars on the measurements are underestimated), according to a $\chi^2$ goodness-of-fit analysis \citep{qin2018bulk, qin2021cosmic, howlett2022sloan}. Thus our investigation aims to test the performance of the Kaiser MLE on mock data to gain a better understanding of its precision and accuracy. We likewise apply the same investigation to the Minimum Variance method (MVE) by \cite{watkins2009consistently}. This estimator has been used in the literature and associated with claims of bulk flows measurements in tension with $\Lambda$CDM. While \cite{agarwal2012testing} has previously shown the estimator to be unbiased and precise for mocks designed for the SFI++, COMPOSITE and DEEP peculiar velocity surveys, we build on this work and further test its performance on current, much larger, datasets. Finally, we also investigate the performance of the MLE method presented in \cite{nusser2014inconsistency} and the Minimum Variance Estimator presented in \cite{peery2018easily}, that are both variants on the Kaiser MLE and Watkins MVE of respectively, and are designed to more accurately capture the bulk flow moment from data, in addition to measure a moment of the data that more closely aligns with the bulk flow as define in Equation~\ref{eq::definition_bulkfflow}. 

In general, bulk flow estimators such as the Kaiser MLE do not obtain the same moment as others such as for example, the Watkins MVE \citep{nusser2014inconsistency, nusser2016methods}. It is important to distinguish that the moments that are estimated may differ and that one should not assume the bulk flow moment is always consistent with Equation~\ref{eq::definition_bulkfflow}. The moments from individual estimators should thus carefully be compared to a theoretical bulk flow amplitude or expectation from simulations, as discussed in \cite{andersen2016cosmology}. 

\subsection{Discussion of methods}

\subsubsection{Kaiser Maximum Likelihood method} 

The Kaiser Maximum Likelihood estimator \citep[Kaiser MLE; ][]{kaiser1988theoretical} is an analytic weighting scheme to derive bulk flow measurements. In this method, the likelihood function for measuring a bulk flow vector $\mathbf{B}$ given a set of $N$ observed radial velocities $u_m$ is written as
\begin{equation}
    \mathcal{L}(\mathrm{B}, u_m) = \Pi_{m = 1}^{N} \frac{1}{\sqrt{2 \pi(\sigma_*^2 + \sigma_m^2)}} e^{-\frac{(u_{m} - \mathbf{B} \cdot \mathbf{\hat{r}}_{m} )^2}{2(\sigma_*^2 + \sigma_m^2)}},
\end{equation}
where $\sigma_m$ is the uncertainty in each velocity measurement, and $\sigma_*$ is an extra component of uncertainty to account for random non-linear motions. If a fixed value for $\sigma_*$ is assumed one can solve for the weights for the galaxy velocities that maximise this likelihood function with respect to the bulk flow by solving $\frac{d\mathcal{L}}{d B_{i}} = 0$. One estimates the bulk flow components along each coordinate axis as 
\begin{equation}
    \tilde{B}_i = \sum_{m=1}^{N} w_{i,m} u_{m},
\end{equation}
and the solution for the weights $w_{i,m}$ are given by
\begin{equation}
    w_{i,m} = \sum_{j} A_{i,j}^{-1} \frac{\hat{n}_{j,m}}{(\sigma_*^2 + \sigma_m^2)},
\end{equation}
where we define
\begin{equation}
    A_{i,j} = \sum_{m} \frac{\hat{n}_{i,m} \hat{n}_{j,m}}{(\sigma_*^2 + \sigma_m^2)}.
\end{equation}
The form of the likelihood function for this method encodes a number of assumptions:
\begin{itemize}
    \item the galaxy PVs are uncorrelated;
    \item the galaxy PVs have errors that are drawn from a Gaussian distribution;
    \item each PV has a component of velocity due to a bulk flow vector that is constant across the entire volume, such that each radial PV can be written as $u_m = B_{\alpha} \hat{n}_{i,\alpha} + \delta_m$, where $\delta_m$ represents a random component of velocity that is not due to the underlying bulk flow \citep{nusser2014inconsistency}; 
    \item that $\sigma_*$ is a fixed value; 
    \item that the velocity field is well modelled by only the bulk flow modes (higher order modes of the field can be neglected);
    \item and that the galaxy PV observations are insensitive to small scale flows.
\end{itemize}

None of these assumptions are generally true. We are interested in exploring how much these may impact the performance of this estimator. 
This may allow us to understand why, when this estimator is applied to realistic mocks, the analysis of the recovered mock bulk flows compared to the true bulk flow moment of each mock generally results in a reduced $\chi^2 > 1$ for the goodness--of--fit \citep{qin2018bulk, qin2021cosmic, howlett2022sloan}, which implies the model for this estimator results in underestimated uncertainties even if the results are unbiased.

In the work of \citet{qin2018bulk}, a modified version of the Kaiser estimator is developed, the $\eta$--MLE estimator. This estimator searches for the Maximum Likelihood bulk flow using a Likelihood function for the observed log--distance ratios $\eta$ of the data rather than the PVs in order to avoid the issue of non--Gaussian uncertainties on PV measurements (as an alternative to the other approaches to deal with this issue, discussed previously). Furthermore they take an MCMC approach to search for the best fitting bulk flow modes and also allow $\sigma_*$ to vary rather than setting a fixed value. Furthermore, in \citet{qin2021cosmic}, this approach is used while also allowing the shear moments of the velocity field to be modelled. In these works the reduced $\chi^2$ is generally still greater than unity. This would imply the assumptions the Kaiser estimator encodes regarding $\sigma_*$, the nature of the PV errors, and the higher order moments of the field (which are no longer assumptions for the $\eta$--MLE method), might be unrelated to issues regarding the precision of the recovered bulk flows using the Kaiser MLE approach. To explore further we investigate the Kaiser estimator in more detail by testing its performance with mocks in section~\ref{sec::mock_performance_eval}. 

\subsubsection{Nusser Maximum Likelihood Estimator}

The Maximum Likelihood method proposed in \cite{nusser2014inconsistency} (hereon Nusser MLE) is very similar to the Kaiser MLE approach. \cite{nusser2014inconsistency} shows the Kaiser MLE method only obtains the bulk flow of a survey volume as defined in Equation~\ref{eq::definition_bulkfflow} if one can reliably assume that the bulk flow is effectively a constant across the survey volume, which is not generally consistent with the definition of the bulk flow as an average of the peculiar velocities in the volume as in Equation~\ref{eq::definition_bulkfflow}. 

In summary, \cite{nusser2014inconsistency} shows in the continuous limit and when there is no angular selection function applied to the galaxies, the estimated quantity for each bulk flow mode obtained by maximising the likelihood function given by the Kaiser MLE approach can be written as 
\begin{equation}
    \tilde{B}_{i} = \frac{3 \int r'^2 dr' d\Omega \frac{w(r') \bar{n}(r')}{\sigma^2(r')} 
 u(\mathbf{r}') \hat{n}(\mathbf{r}')}{4 \pi \int r'^2 dr' \frac{w(r') \bar{n}(r')}{\sigma^2(r')} }.  
\end{equation}
In this equation $\Omega$ is a solid angle, $\sigma$ is the uncertainty on the galaxy PV (which in this limit is dependent only on $r$) and $\bar{n}$ is the mean number density of galaxies (which is also assumed to be dependent only on $r$). For the Kaiser MLE estimator we have $w(r) = 1$. \cite{nusser2014inconsistency} shows that this equation is only consistent with the definition of the bulk flow in Equation~\ref{eq::definition_bulkfflow} if one sets $w(r) = \frac{\sigma^2(r)}{n(r) r^2}$. In the Nusser MLE, this modification has been adopted to re--derive the analytical solution to the maximum likelihood weights. In the case that PV field is curl--free and the radial selection function is spherically symmetric, the \cite{nusser2014inconsistency} modification results in the following weighting scheme,
\begin{equation}
    w_{i,m} = \sum_{j} A_{i,j}^{-1} \frac{\hat{n}_{j,m}}{(\bar{n}_m d_m^2 )},
\end{equation}
where we define
\begin{equation}
    A_{i,j} = \sum_{m} \frac{\hat{n}_{i,m} \hat{n}_{j,m}}{(\bar{n}_m d_m^2 )}.
\end{equation}
In the above equations $\bar{n}_m$ is the number density of galaxies at the comoving distance $d_m$ the galaxy appears to be from the observer. \cite{nusser2014inconsistency} shows this weighting scheme obtains a more accurate bulk flow estimate as defined in Equation~\ref{eq::definition_bulkfflow} than the Kaiser MLE weighting scheme, using a single simulation. This method has the advantage over the Kaiser method of not requiring the bulk flow to be well-represented as a constant--valued vector across the survey volume, although it has the disadvantage of requiring spherical symmetry in the survey geometry, and otherwise shares the same assumptions that are encoded in the Kaiser method. We investigate its performance on mock data here. 

\subsubsection{Watkins, Feldman and Hudson Minimum Variance method}

For comparison we are also interested in exploring the measured bulk flows obtained from mocks when applying the Minimum Variance estimator (MVE) method by \cite{watkins2009consistently}. This estimator was shown to be unbiased in \citet{agarwal2012testing} however we are interested in also exploring the precision and accuracy of this estimator in a similar way to the Kaiser MLE method.

The aim of the MVE method is to obtain a bulk flow measurement from the data that minimizes the variance in the difference between the bulk flow estimate $\tilde{B}_i$ of the data and the measured bulk flow that would be obtained from a survey with an ideal window function, $U_i$. This in principal allows for the measured bulk flow to be more comparable to other surveys in which this method has been applied even when they have different geometries, as long as the window function of $U_i$ is the same. The approach involves the use of the Lagrange multiplier method in order to also satisfy a constraint so that the measured bulk flow amplitude $\tilde{B}_i$ is correct on average. The Lagrangian function to be minimized with respect to the weights $w_{i,m}$ for each $m^{\mathrm{th}}$ galaxy contribution to the $i^{\mathrm{th}}$ bulk flow mode is given by
\begin{equation}
    \mathcal{L} = \langle (\tilde{B}_i - U_i)^2 \rangle + \lambda_{ij} \sum_m w_{i,m} \hat{n}_{i,m},
\end{equation}
where $\lambda_{ij}$ is the Lagrange multiplier and $\hat{n}_{i,m}$ is the same as previously. The resulting weighting scheme can be found after solving for $\frac{d\mathcal{L}}{dw_{i,m}} = 0$. More details can be found in \cite{watkins2009consistently, feldman2010cosmic, agarwal2012testing, scrimgeour20166df}. Like the Kaiser method, this method also incorporates, although much more weakly, the assumption that the measured radial PVs of galaxies contain a component that is due to a constant bulk flow vector across the entire survey volume; in fact \cite{nusser2016methods} shows that these methods are equivalent in the limiting scenarios that the correlations between the ideal survey galaxies and real survey galaxies approaches zero. 

We can expect some of the assumptions listed previously which may affect the results from the Kaiser MLE approach (that $\sigma_*$ can be a fixed value, that the galaxy PV measurements are drawn from a Gaussian PDF) also affect the results from the Watkins MVE in a similar way \citep[although as discussed previously, there are approaches in the literature to Gaussianise PV measurements given in][]{watkins2015unbiased, qin2021gaussianization, hoffman2021cosmicflows}. However, the Watkins MVE does take into account that there are non--zero linear correlations between galaxy peculiar velocities. It is also less sensitive to small scale flows because of the constraint it enforces on the window function of the measured bulk flow moments. Both this method and the Kaiser method suffer from the potential risk due of leakage of signal due to higher order modes in the velocity field affecting the bulk flow estimate, although in \cite{feldman2010cosmic} this method is extended to include modelling for higher order moments of the field. On the other hand, in \cite{feldman2010cosmic} they find that not including the higher order moments of the field such as the shear and octupole did not contaminate their bulk flow measurements. In our own tests with the Kaiser MLE in which we extended the method to allow modelling for higher order modes and applied it to realistic mocks for the SDSS data, we found that failing to model the higher order modes did not reduce the accuracy of the bulk flow measurements. This is discussed further in section~\ref{sec::results}.

\subsubsection{Peery Minimum Variance Estimator}

Finally we discuss the Minimum Variance Estimator that is presented in \cite{peery2018easily} (hereon, Peery MVE). This is effectively the same as the MVE method developed by \cite{watkins2009consistently} discussed previously, but with two modifications. Firstly, the ideal survey that is used to constrain the window function of the estimated bulk flow in the MVE method is modified to follow a radial distribution with selection function such that the number density of objects $n(r)$ is proportional to $r^{-2}$, where $r$ is the radial distance to objects in the ideal survey. Or alternatively, the objects in the ideal survey follow a uniform radial selection function, but are weighted by an additional factor of $r^{-2}$ compared to the ideal weights in the original MVE method. By following either of these approaches this ensures that one obtains an estimate of the bulk flow much more closely aligned to the moment defined in Equation~\ref{eq::definition_bulkfflow} without encoding the assumption that the bulk flow is constant across the survey volume, as is true in the Nusser MLE method. The authors arrive at this scheme following and expanding on the derivations shown in \cite{nusser2014inconsistency} for application to the MVE scheme. 

Secondly, an additional constraint equation is introduced to the Lagrangian function, that ensures $\sum_{m} cz_{m} = 0$. This constraint is introduced in order to allow the estimator to be independent on uncertainty in the Hubble constant $H_0$. Effectively, we can understand that this constraint equation should make the estimated bulk flow independent of the global zero--point calibration of the dataset. More details regarding this estimator can be found in \cite{peery2018easily}. We expect that this variant of the MVE method thus has two advantages over the Watkins MVE approach, and unlike the Nusser MLE approach does not encode any assumptions about the survey geometry. We present results for the application of this estimator to mock data also.  

\section{Performance evaluation: bulk flow estimators}\label{sec::mock_performance_eval}

In this section, we use different sets of mock PV survey data to test the performance of the estimators discussed previously to accurately and precisely recover the bulk flow. For all mocks, we place galaxies at their observed redshifts. We begin with results from tests on simplistic mocks before moving on to more sophisticated mock data that involves modelling for the growth of structure in the Universe. We conclude by testing these estimators on the fully realistic mocks from numerical simulations that were described in section~\ref{sec::data_and_realisticmocks} and use these results to comment on the performance of the estimators to recover the bulk flow in real data. 

It should be noted, the simplistic mocks and Zeldovich mocks (described in the following sections) were produced for analyzing the performance of the bulk flow estimators specifically when the survey geometry is altered or the statistical properties of the velocity distribution is changed, without the additional complexities that are introduced in realistic mocks for data. As stated previously, the goal has been to thoroughly evaluate the estimators and determine what systematics may affect their performance; the simplistic mocks and Zeldovich mocks are, unlike the realistic mocks for CF4, not representative of any realistic datasets but are used to gain a general sense of the limitations of the estimators. For reasons that will become apparent, we also do not apply all four estimators for every test or set of mocks with different properties that we construct.  

\subsection{Tests on simplistic mocks}

\subsubsection{Generating the mocks}

The simplistic mocks were generated as follows: 

\begin{itemize}
    \item We generate points within a sphere with a chosen radial distribution representing galaxies (various are used) and record right ascension (RA), declination (Dec), cosmological redshift ($z_{\text{rec}}$), and the radial comoving distance to the point from the observer in real space ($D(z_{\text{rec}})$). To obtain a radial distribution of objects that is not uniform, we simply use downsampling to obtain the desired distribution of points as a function of comoving distance.  
    
    \item We generate a radial velocity covariance matrix for the data points using linear theory (see Appendix~\ref{appendix:lineartheorycovariance} for more detail). The covariance matrix depends on a chosen cosmological model (given in the next section) and the coordinates for the data points. We use the linear theory covariance matrix to draw Gaussian random velocities for each object in the mock.  
    
    \item We generate a random bulk flow vector with an $x$, $y$ and $z$ component by drawing from a uniform distribution for each mode. We treat this bulk flow vector as the true bulk flow of the mock data we expect our estimator to recover. The radial component of this bulk flow is added to the radial PVs drawn in the previous step, which are then treated as `true' radial PVs of each galaxy. Each galaxy thus has a velocity component that is due to the same underlying bulk flow $\mathbf{B}$ and their own random component of velocity due to statistical variation generated from linear theory in the previous step, $u_{\text{linear,m}}$. Overall for each object labelled by $m$ we have 
    \begin{align}
        u_{\text{total},m} & = u_{\text{linear},m} + \mathbf{B} \cdot \hat{\mathbf{r}}_m \\ \nonumber
        & = u_{\text{linear},m} + B_x \cos(\theta_m) \sin(\phi_m) + B_y \sin(\theta_m) \sin(\phi_m) \\ \nonumber 
        & + B_z \cos(\phi_m).
    \end{align} 
    where $\theta$ and $\phi$ are angles in spherical polar coordinates that can be mapped to RA and Dec for each object. 
    
    \item The radial PVs from the previous step are used to determine a redshift observation in the CMB--frame for each object. We use the equation
    \begin{equation}
        z_{\text{CMB, obs}} = (1 + z_{\text{rec}})(1 + z_{\text{pec}}) - 1.
    \end{equation} 
    For each object we already have $z_{\text{rec}}$ from the first step. $z_{\text{pec}}$ (the Doppler redshift due to the radial PV of the galaxy) can be calculated from the radial PVs generated in the previous two steps. We use the special relativistic Doppler shift equation.
    
    \item Finally we use $z_{\text{CMB, obs}}$ and $z_{\text{rec}}$ to compute truth log--distance ratios $\eta_t$ for each galaxy as $\eta_t = \log_{10}{\left(\frac{D(z_{\text{CMB}})}{D(z_{\text{rec}})}\right)}$. To obtain observed log--distance ratios $\eta_o$, we draw random $\eta_o$ for each galaxy from a Gaussian probability distribution function with a mean given by $\eta_t$ and a standard deviation $\sigma_{\eta}$ given by a constant. This constant is used to represent the uncertainty in the observed $\eta_o$. We choose $\sigma_{\eta} = 0.05$. 
    \item Finally we use the estimator of \cite{watkins2015unbiased} to convert the log--distance ratio observations and their uncertainties to radial PVs and PV uncertainties for each object,
    \begin{equation}
        u_m \approx \frac{c z_{\text{mod}}}{1 + z_{\text{mod}}} \eta_0 \ln(10)
    \end{equation}
    and 
    \begin{align}
        z_{\text{mod}} = z_{\text{CMB}} [1 + \frac{1}{2}(1 - q_0)z_{\text{CMB}} 
        - \frac{1}{6}(j_0 - q_0 - 3 q_0^2 + 1)z_{\text{CMB}}^2],
    \end{align}
    where $q_0$ and $j_0$ are the deceleration and jerk parameters, respectively. 
\end{itemize}

\subsubsection{Results summary: simplistic mocks}

\begin{figure*}
    \centering
    \includegraphics[scale=0.47]{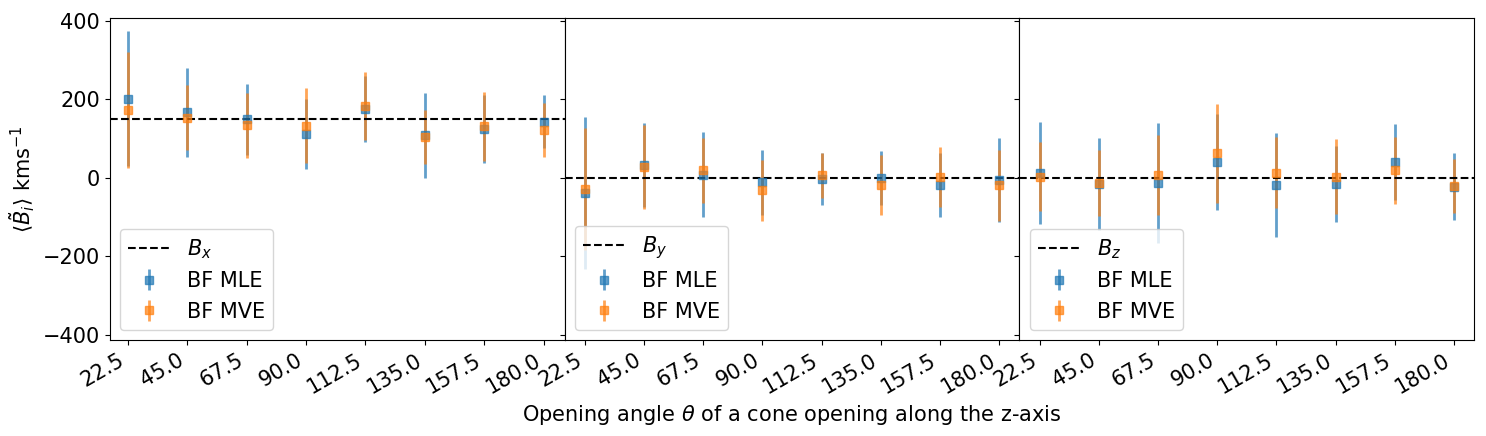}
    \caption{The averaged recovered bulk flow $\langle \tilde{B}_i \rangle$ component from simplistic mocks with a cone geometry. Each panel shows $\langle \tilde{B}_i \rangle$ in the $x$, $y$ and $z$ directions. The value for $\mathbf{B}$ (the underlying constant-valued bulk flow vector) is the same for each mock and the components are shown by the horizontal dashed line in each panel. Each point shows the average recovered bulk flows of 32 mocks with the same opening angle $\theta$, plotted against $\theta$; $\theta = 180$ corresponds to a spherical survey, $\theta = 90$ corresponds to a hemisphere. For all the mocks the spatial distribution of objects follows a Gaussian radial distribution such that $n(r) \propto e^{-r^2}$ with a standard deviation of $\sim 50 \mathrm{Mpc h}^{-1}$. The recovered bulk flows from the Kaiser MLE and the Watkins MVE methods are shown in blue and orange respectively. The error bars on each point represent the standard deviation of $\langle \tilde{B}_i \rangle$. }
\label{fig:bulkflow_cone_geometries_varying}
\end{figure*}

\begin{figure}
    \centering
    \includegraphics[width=0.48\textwidth]{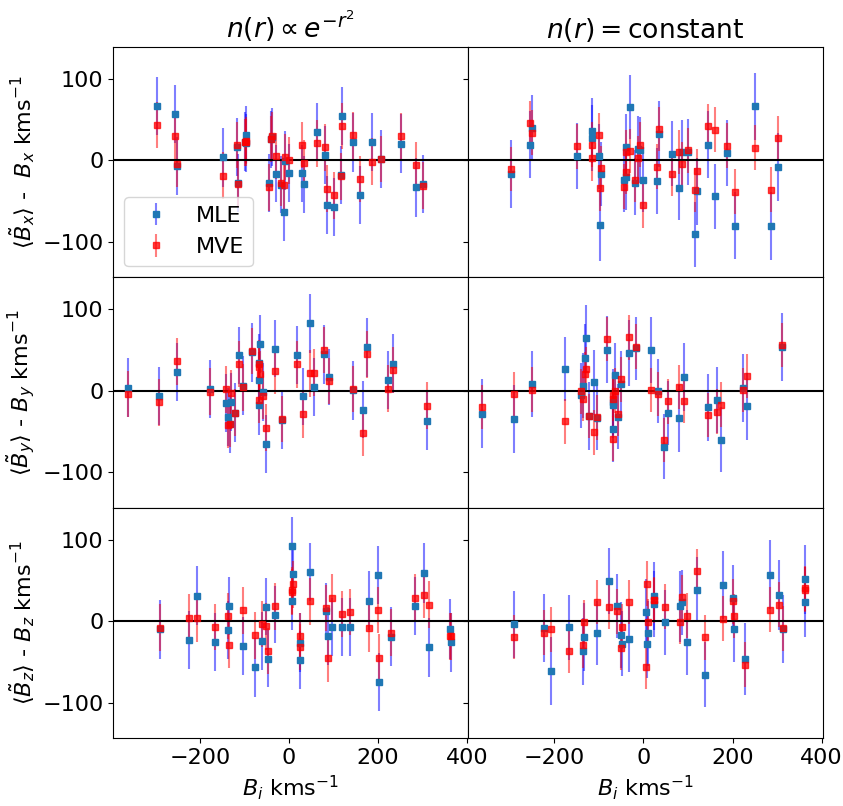}
    \caption{
    The difference between the recovered bulk flow components $\langle \tilde{B}_i \rangle$ from the components of the constant-valued bulk flow vector $\mathbf{B}$ for 512 mocks using the Kaiser MLE (blue points) and Watkins MVE (red points), plotted against $B_i$ (the components of $\mathbf{B}$) for simplistic mocks. The panels from top to bottom show the results for the $x$, $y$ and $z$ directions. The results in the left panel are for a set of mocks with a  selection function given by a Gaussian radial distribution, the right column of panels a uniform radial distribution. Each data point is the average recovered bulk flow from 32 mocks with the same vector $\mathbf{B}$ added to the data. The error bars give the standard deviation of $\langle \tilde{B}_i \rangle$. }
\label{fig:bulkflow_selectionfunction_varying}
\end{figure}

Simplistic mocks are generated each with 5000 galaxy PVs. We choose a $\Lambda\text{CDM}$ cosmological model with $H_0 = 69 \mathrm{km s}^{-1} \mathrm{Mpc}^{-1}$, $\Omega_m = 0.31$, $\Omega_{\Lambda} = 0.69$. We generate sets of the mocks with varying geometries by 1) altering the radial selection function by downsampling to a desired number density of objects from an uniform distribution, as discussed previously, and 2) generating objects in a desired sky mask (sky coverage) by limiting the allowed values for $\theta$ and $\phi$ for each object. In summary, we found the Watkins MVE and Kaiser MLE were able to generally obtain an unbiased estimate $\Tilde{\mathbf{B}}$ of the bulk flow vector $\mathbf{B}$ added to the mocks, regardless of the radial selection function of the points or the sky mask applied to the data. In general we find, the reduced $\chi^2$ goodness--of--fit of the recovered bulk flows to the true underlying bulk flows obtains $\chi^2 \sim 1$. These conclusions are demonstrated in Figure~\ref{fig:bulkflow_cone_geometries_varying} and Figure~\ref{fig:bulkflow_selectionfunction_varying}. It is also apparent from these tests, that when the bulk flow vector can be described as a constant valued vector across the survey volume, then the bulk flow moments obtained from these different estimators are actually consistent with each other and also the bulk flow moment as defined in Equation~\ref{eq::definition_bulkfflow} \citep{nusser2016methods}.

Figure~\ref{fig:bulkflow_cone_geometries_varying} shows the results of applying the Watkins MVE and Kaiser MLE method to mocks with a set bulk flow $\mathbf{B}$ with the shape of a cone with varying opening angles ranging from a narrow cone to a full spherical mock, and with the number density of objects $n(r)$ at a given radius from the observer, $r$, given such that $n(r) \propto e^{-r^2}$. Figure~\ref{fig:bulkflow_selectionfunction_varying}
shows the results of applying both estimators to sets of fully spherical mocks with two different radial selection functions specified just above the plot panels and with various bulk flows. In all cases $\Tilde{\mathbf{B}}$ from both of the estimators is unbiased compared to $\mathbf{B}$ and $\chi^2 \sim 1$. This tells us that overall the Watkins MVE and Kaiser MLE are good estimators for $\mathbf{B}$ in this simplistic model where the radial peculiar velocities have a component due to $\mathbf{B} \cdot \hat{\mathbf{n}}$ and a Gaussian random component of motion. In this model, the survey geometry does not affect the ability of these estimators to recover $\mathbf{B}$ on average. More supporting results for these mocks not included here for conciseness can be found \hyperlink{https://github.com/abbew25/Measuring\_bulkflows}{here}. Given the nature of these mocks, we can expect the results will not differ for the Peery MVE estimator and do not show results when applying these estimators; however we will see the Nusser estimator may indeed suffer when a non-spherical sky mask is used. 

\subsection{Tests on mocks with linear structure growth}

\subsubsection{Generating the mocks}

The second set of mocks we consider for testing are more realistic in that they include structure growth in the Universe which changes the spatial distribution of objects compared to the simplistic mocks described previously. However, these have been deliberately made to incorporate only linear structure growth of matter as closely as possible. We refer to these mocks as `Zeldovich mocks'. We generate these with the same cosmological model as specified for the simplistic mocks in the previous section. 

We use the $n$--body simulation code L-PICOLA \citep{howlett2015picola} to generate these mocks. The L-PICOLA code uses an approximate analytical solution to calculate the initial velocities and positions of dark matter particles due to linear structure growth, called the Zeldovich approximation. It then evolves the simulation with a numerical algorithm to accurately capture the effects of non--linear structure growth from a specified initial redshift by the user to a final redshift. We expect that non--linear structure growth has an effect on the PVs in the simulation because it is known that the galaxy PVs follow a distribution with higher kurtosis (or extended tails) compared to a Gaussian distribution when non--linear structure growth is present \citep{sheth2001peculiar}. We hypothesised that the altered velocity distribution in the mocks when non--linear structure growth is present, in addition to the complexity that is introduced into the spatial distribution by structure growth, has an effect on how well bulk flow estimators recover the bulk flow. Therefore to isolate the effect of altering the velocity distribution due to non--linear structure growth and test how well the estimators work when it is not present, we evolve only the linear structure growth (which is calculated with the Zeldovich approximation) in the $n$--body code and factor out the evolution that causes non--linear structure growth. This can be done using L-PICOLA by setting the initial and final redshift in the simulation both to $z = 0$. Since there is only linear structure growth, we do not identify dark matter halos and or galaxy populations within them, but instead assume that each dark matter particle in the simulation can be approximated as a single galaxy. This is a valid choice for our purposes given we expect the PV field to be largely insensitive to galaxy bias \citep{zheng2015determination}. For each simulation box we create four mocks by specifying four different origins for an observer within the box, and we ensure the distance between the four origins is enough such that data from the mocks is not overlapping and that the measured data will not be correlated. 

The realistic mocks in \cite{qin2018bulk, qin2021cosmic, howlett2022sloan} are created using the L-PICOLA code but \emph{do} include non--linear structure growth. We can compare how well the Kaiser MLE measures the bulk flow using the Zeldovich mocks described here and compare the results to those found by \cite{qin2018bulk, qin2021cosmic, howlett2022sloan} who apply the Kaiser MLE (or variations of it) to their mocks to determine whether there is any relation between the performance of these estimators and the shape of the distribution of PVs. Therefore, for some of our tests we create SDSS PV survey simulations with the Zeldovich mocks described here by following the exact procedure implemented by \cite{howlett2022sloan} to create the SDSS mocks, aside from the main difference being that L--PICOLA full--numerical simulations are not used to generate galaxy positions and velocities. For tests with these mocks we focus on just the Kaiser MLE approach since our goal is to purely isolate the effect of having non--linear structure growth. 

We apply tests with all four of the estimators described previously to more general Zeldovich mocks. For these more general mocks, with are fully spherical and contain just 5000 galaxies (by downsampling from simulations) with various different choices of radial survey selection functions, the log--distance and PV observations are simulated in the same way as the steps described for the simplistic mocks. We generate observations of $\eta$ from the true values of $\eta$ by drawing Gaussian random values with a set value for $\sigma_{\eta}$ for the Gaussian (which becomes the uncertainty on the observation). The estimator by \cite{watkins2015unbiased} is also used to convert the log--distance ratios to PV observations. With these mocks, we can also test whether other factors that are present in the mocks, such as the properties of the spatial and velocity distribution or survey geometry affects the precision or accuracy of bulk flow estimators, while having factored out further complexity due to non--linear structure growth.

\subsubsection{Results summary: Zeldovich mocks}

We begin by noting here that we compare the estimated moment $\tilde{\mathbf{B}}$ from the Peery MVE (that we will hereon label $\tilde{\mathbf{B}}_{\mathrm{P}}$) and the Nusser MLE ($\tilde{\mathbf{B}}_{\mathrm{N}}$) to the moments $\mathbf{B}_{\mathrm{P}}$, $\mathbf{B}_{\mathrm{N}}$ respectively, defined as the sum of the true 3D PVs of each object in the simulation (essentially Equation~\ref{eq::definition_bulkfflow} for both estimators). However, for the Kaiser MLE and the Watkins MVE the estimated moment slightly differs, as mentioned previously. For the Kaiser MLE, the true 3D PVs of each galaxies are weighted by $(\sigma_n^2 + \sigma_*^2)^{-1}$ to define $\mathbf{B}_{\mathrm{K}}$ and for the Watkins MVE by $e^{-r^2/(2R^2)}$ to define $\mathbf{B}_{\mathrm{W}}$, where $R$ is the standard deviation of the Gaussian used to define the window function of the ideal survey used in this estimator. This choice is taken because 
the Watkins MVE attempts to obtain a bulk flow estimate from the (ideal) Gaussian--weighted volume, and the Kaiser MLE obtains a bulk flow estimate of the most likely constant bulk flow vector across the volume, with the velocities weighted by the uncertainty on each observation. \footnote{One might expect based on how the MVE estimator is designed to weight the galaxies such that the bulk flow window function matches a chosen ideal survey, that we should actually compute the expected bulk flow amplitude as the average of all the velocities in the simulation within the volume defined by the ideal radius. However we found in general that this moment is not well-captured by the estimator and the calculated moment is more accurately and precisely captured by the moment we define for comparison in our tests. }

We summarise our findings from a number of tests performed on the Zeldovich mocks here, while more evidence and plots to support these results can be found \hyperlink{https://github.com/abbew25/Measuring\_bulkflows}{here}.

Firstly, we tested the Kaiser MLE on the SDSS--Zeldovich mocks.  We found that when applying the Kaiser MLE to these mocks, the results are unbiased, similarly to results found in \cite{howlett2022sloan}. However, we also find that the precision of the recovered bulk flows is under-reported; the reduced $\chi^2$ is much greater than unity, similar to that found by \cite{howlett2022sloan} with the L--PICOLA mocks. Therefore we conclude from this result that the effects of non--linear structure growth on the spatial or velocity distribution of galaxies is not related to the precision or under--reporting of errors for the Kaiser MLE method in general. We also expect that we would find similar results using other estimators given that they do not include any additional modelling for non--linear theory PVs. 

We also extended the analysis for the SDSS-Zeldovich PV mocks to incorporate modelling for higher order modes of the PV field \citep[see the definition of the shear modes in section~3 of][and references within]{qin2019redshift} by simply substituting the expansion of the PV field with the mode expansion into the likelihood function for the Kaiser MLE approach, in place of the field expansion defined as the bulk flow modes only, $\mathbf{B}\cdot\mathbf{\hat{r}}$. We found that the resulting bulk flows were unbiased but the extra degrees of freedom did not improve the precision of the recovered bulk flows. We thus conclude aliasing due to a failure to model higher order modes of the velocity field is not influencing how precisely the Kaiser MLE is estimating the bulk flow with this set of mocks. We do not extend this test to the Watkins MVE as this result has effectively been shown already to be true by \cite{feldman2010cosmic} for this estimator, although we might expect the importance of modelling these modes to be more largely dependent on the data or mocks used. 

In further tests on more general Zeldovich mocks with 5000 objects and varying choices for the radial selection function of the mock data, we tested all four estimators. In general we find when applying the Kaiser MLE or Watkins MVE to these mocks, in all choices for the radial selection function of the survey the estimated bulk flows are unbiased compared to the expected moments and both estimators in fact give consistent results to each other for the estimated moments. However, the reduced $\chi^2$ is usually greater than unity. Additionally for both methods, the reduced $\chi^2$ changes depending on the radial selection that is applied to the mock catalogues, in contradiction to our results with the simplistic mocks. We find there is a correlation between the $\chi^2$ goodness of fit and properties related to the radial selection function applied to the data. 

In contrast, when we apply the Peery MVE and Nusser MLE, which also obtain unbiased estimates of the bulk flow, there is no correlation present when applying them to the same sets of mocks, and in general the reduced $\chi^2$ is closer to unity, regardless of the radial selection function. This would lead us to conclude, that an underlying assumption that plays a role is the low--precision in results is that the bulk flow vector cannot be well represented as a constant--valued vector across the survey volume and that the radial selection function of the data may need to be taken into account for the bulk flow measurement; this is one of the differences between the Watkins MVE and Kaiser MLE compared to the Peery MVE and Nusser MLE.

\begin{figure}
    \centering
    \includegraphics[width=0.485\textwidth]{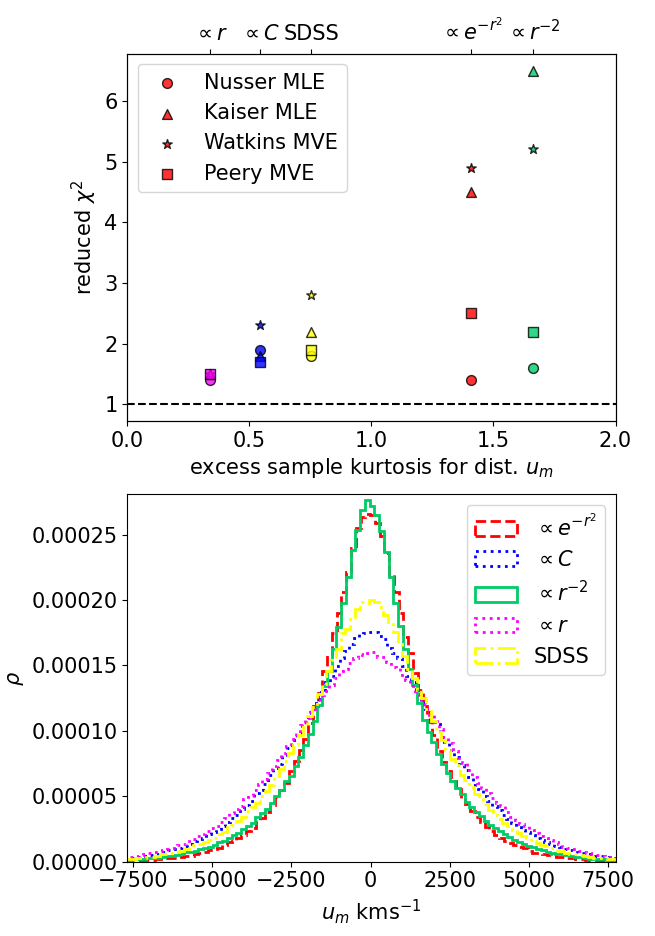}
    \caption{
    Top panel: observed trend between the reduced $\chi^2$ squared goodness--of--fit of recovered bulk flows to the expected moments of the bulk flows from a set of the Zeldovich mocks with a spherical sky mask, plotted against the kurtosis of the distribution of observed velocities for the set of mocks. Points with the same colour define results for estimators applied to the same set of mocks with a radial selection function that is proportional to a number density of objects specified by the labels on the $x$--ticks on the top of the plot. Different shaped markers distinguish between results from applying different estimators, as given in the legend. Lower panel: density plot of the distributions of observed PVs for mocks with different choices of radial selection function.
}\label{fig:trend_kurtosis_and_chisquare_kaiser_nusser_zeldovichmocks}
\end{figure}

Regarding the correlation we observe between the radial selection function of the survey and $\chi^2$ goodness--of--fit, we note that when the observed log--distance ratios $\eta_o$ for the mock data are drawn from Gaussian distributions centered on $\eta_t$ for each object with a constant $\sigma_{\eta}$, the distribution of observed PVs estimated following the estimator by \cite{watkins2015unbiased} changes in shape and kurtosis depending on the radial selection function of the data, although the distribution of true PVs from the mocks does not. We emphasise here that this is \emph{not} a bias or flaw concerning this PV estimator. However the changing in the PV distribution can be understood because the spread and tails (which we quantify using the excess sample kurtosis of the distribution compared to a Gaussian distribution) of the distribution of \emph{true} galaxy log--distance ratios (with a mean of zero) is expected to be smaller when there are more objects at a further distance from the observer where we might expect the ratio $\frac{D(z_{\text{CMB}})}{D(z_{\text{rec}})}$ to be closer to one. We then expect relative to this, the distribution of \emph{observed} galaxy log--distance ratios will have greater spread. Thus for a choice of a constant $\sigma_{\eta}$ in our mocks, changing the radial selection function of the mock data will change the observed distribution of log-distance ratios due to the way $\frac{D(z_{\text{CMB}})}{D(z_{\text{rec}})}$ is affected and the same applies to the distribution of observed PVs that are estimated from the log-distance ratios. Therefore overall, we might understand that depending on the radial selection function of the data, which changes the average radial distance to objects (for a fixed number of 5000 galaxies) the kurtosis of the distribution of $\eta_o$ and likewise the PVs is altered. As the percentage error on each $\eta_o$ must be more significant relative to their true value $\eta_t$ when the distribution of $\eta_t$ changes for a fixed choice $\sigma_{\eta}$, we see some correlation between the kurtosis of the PV distribution (which has a shape following the $\eta_o$ distribution when the PV estimator by \cite{watkins2015unbiased} is used) and the $\chi^2$ goodness--of--fit to the measured bulk flows. However we would expect that if the uncertainty in the measured bulk flows are well captured by the choice of weighting from the bulk flow estimator and uncertainties given to each data point, this might not be the case. The excess sample kurtosis of the resulting PV distribution appears to be correlated with the reduced $\chi^2$ for a set of mocks that have the same radial selection function, when applying the Kaiser MLE or Watkins MVE. This is demonstrated in Figure~\ref{fig:trend_kurtosis_and_chisquare_kaiser_nusser_zeldovichmocks}. The correlation is not apparent for the results from the Peery MVE or Nusser MLE.

This is interesting, because the motivation behind the Nusser MLE in its derivation was to solve for an improved weighting scheme for the bulk flow of the volume, without requiring the assumption of a constant valued vector across the survey volume. The Peery MVE estimator attempts to achieve the same. Our results suggest that it is this assumption that causes the underestimation of the error bars in the standard Kaiser and Watkins estimators, in a complex way that is also tied to the PV survey geometry. In general, the conclusion is that the Peery MVE and Nusser MLE should be applied for bulk flow estimates, although the Nusser estimator does rely on another assumption regarding spherical symmetry, which the Kaiser MLE approach does not rely on. Unfortunately, as commented on further in the next section we find that this assumption breaks down catastrophically for the CF4 geometry, which makes the Nusser MLE estimator unsuitable.

\subsection{Tests on realistic mocks for CF4 data}

In this section, we test the performance of the bulk flow estimators on mocks which fully capture the same selection effects as the CF4 dataset and non--linear structure growth. The details of the process to generate these mocks were discussed in section~\ref{sec::data_and_realisticmocks}. The cosmological model is the same as for previous mocks. We tested the performance of the Kaiser MLE, Nusser MLE and the Peery MVE in detail, and report primarily on the performance of the first and last of these. This is for the following two reasons:
\begin{enumerate}
    \item {We only test the Peery MVE, rather than also testing the Watkins MVE, due to the conclusions from the previous results but also as we expect it to be more useful given that it incorporates a constraint to ensure the results are independent of systematic errors in the zero--point calibration applied to the CF4 data.}
    \item {We found that the Nusser MLE gives a biased result on the CF4 mocks with extremely large scatter between mock realisations, which we attribute to the lack of spherical symmetry in the geometry of the CF4 data, which is an underlying assumption of the Nusser MLE approach. These results can be seen \hyperlink{https://github.com/abbew25/Measuring\_bulkflows}{here}.}
\end{enumerate} 
For the Kaiser MLE and Peery MVE, we next focus on judging how precisely and accurately these estimators are able to constrain the bulk flow using these methods on the data. 

\subsubsection{Kaiser MLE method: results for CF4 mocks}

\begin{figure*}
    \centering
    \begin{subfigure}{1.0\textwidth}
      \centering
      \includegraphics[width=1.0\linewidth]{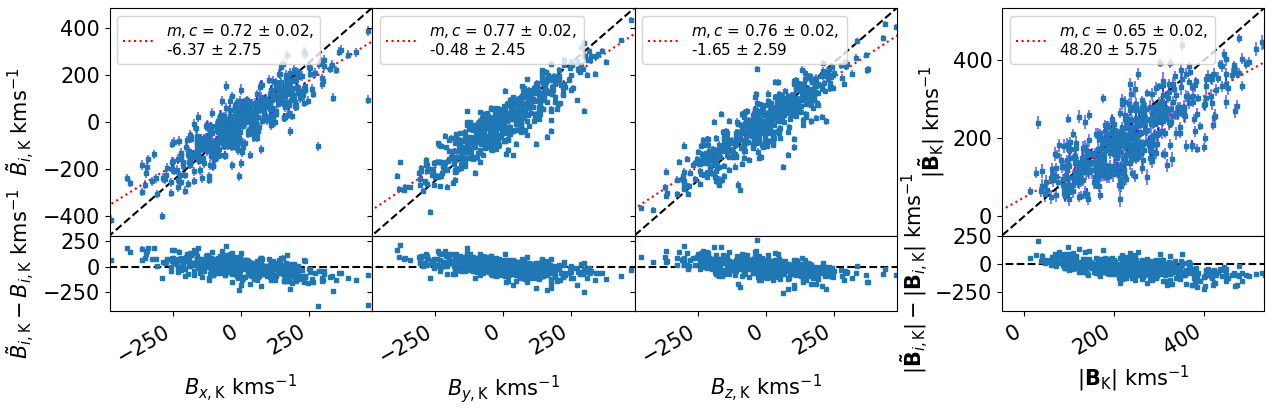}
      \caption{}
      \label{fig:results_kaiser_realisticmocks_CF4}
    \end{subfigure}%
    
    \begin{subfigure}{1.0\textwidth}
      \centering
      \includegraphics[width=1.0\linewidth]{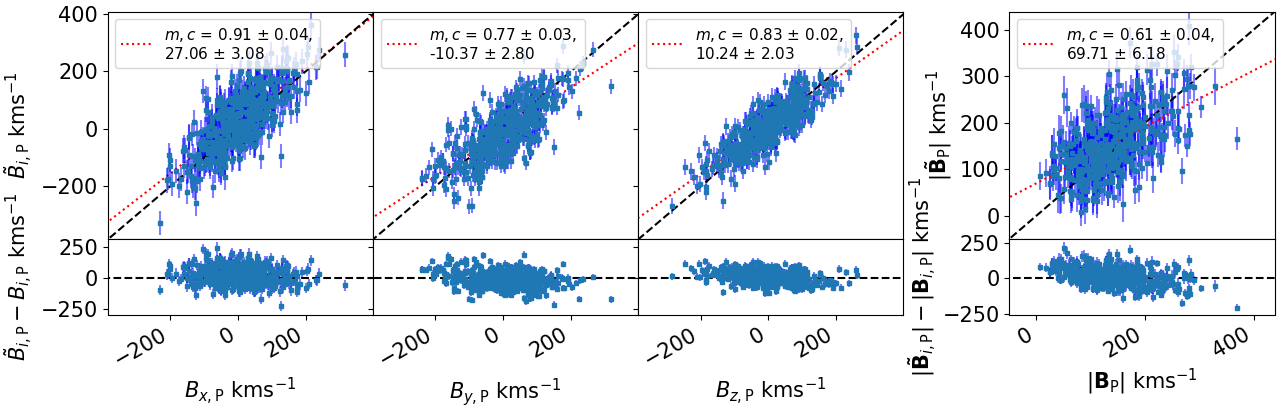}
      \caption{}
      \label{fig:results_peery_realisticmocks_CF4}
    \end{subfigure}
    \caption{a) The recovered bulk flow components $\tilde{B}_{i,\mathrm{K}}$ vs the $B_{i,\mathrm{K}}$ for the realistic mocks for CF4 when applying the Kaiser MLE method, shown in Supergalactic coordinates. The top panels show $B_{i,\mathrm{K}}$ against $\tilde{B}_{i,\mathrm{K}}$ directly while the lower panels plot $B_{i,\mathrm{K}}$ against the residual $\tilde{B}_{i,\mathrm{K}} - B_{i,\mathrm{K}}$. In the top panels the black dashed line is simply a 1--to--1 line and in the lower panel the black dashed line is a horizontal line for comparison. The red dashed line in the top panel is the best fit linear regression to the data in each panel, where $m$ gives the gradient and $c$ gives the y--intercept. b) As in a), but when we apply the Peery MVE method.}
\end{figure*}

We applied the Kaiser MLE to 512 CF4 mocks. We also tested the results when a distance limit in the mock data was introduced at various cut--off scales $r_c$ of between 35 and 345 Mpc $h^{-1}$, such that objects included involve only those with distances $\leq r_c$. This was because for the real CF4 data, we want to investigate the estimated bulk flow when data is included at different scales and this required us to validate the performance of the estimator on mocks with the same distance limits applied. The `truth' bulk flow moment, $\mathbf{B}_{\mathrm{K}}$, which we compare our results to, was modified to incorporate only the PVs of objects not excluded by the cut--off scale when running these tests. In general we found for any cut--off scale or when all the data is included, the estimated bulk flow is strongly correlated to the truth bulk flow. On average however, the bulk flow is slightly underestimated compared to the expected moment of the data, and this is apparent from the slopes of the best fit linear regressions in Figure~\ref{fig:results_kaiser_realisticmocks_CF4} which shows the results from the Kaiser MLE applied to the 512 CF4 mocks. In this plot we have used all the data and there is no cut--off scale applied. We found similar results when applying the estimator to the same mocks, but when using the true radial peculiar velocity of each galaxy with zero uncertainty, instead of using observed PVs calculated from observed log--distance ratios. Therefore this bias is due to the estimator and not due a systematic or error in the observational mock data. The results are shown in plots for the different cut off scales and for the true radial velocities which can be found \hyperlink{https://github.com/abbew25/Measuring\_bulkflows}{here} for the interested reader. 

Given these results, we could consider correcting the bulk flow that we will obtain from the real data with the Kaiser MLE method using the mock results, to ensure we can compare the bulk flow that is recovered to a cosmological model. Furthermore, we can use the results from the CF4 mocks where we have used the true radial PVs of each object (as described above) rather than the observational PVs. By averaging the bulk flow from many CF4 mocks and taking the standard deviation of them, this gives us a good representation of what we should expect for 1) the bulk flow on a given scale and for a given the cosmological model, and 2) for the cosmic variance uncertainty on the bulk flow. By using these for comparison, we are able to take into account not only how geometry should affect the bulk flow estimate but also any systematics that influence the performance of the Kaiser MLE method, while not being affected by uncertainties in PV observations.

\begin{figure}
    \centering
    \includegraphics[width=0.48\textwidth]{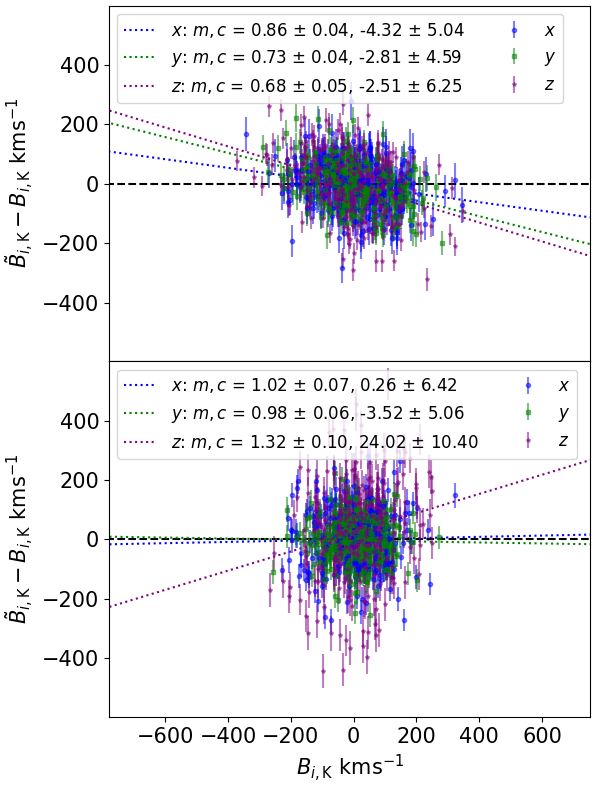}
    \caption{Resulting bulk flow residuals when applying the Kaiser MLE approach to SDSS mock data (256 mocks). The upper panels are mocks that include non--linear structure growth in the PV field, while only Zel'dovich theory for structure growth is included in the PV field for the mocks in the lower panels. The coloured dashed lines show the best fit linear regression in each panel for each coordinate direction, with the legend giving the slope and y-intercept for the regression. The black line gives a horizontal line for comparison. It should be noted here that the mocks do have slightly differently defined projections onto the $x$, $y$ and $z$ axes for the observer which may affect the scatter seen in the measured bulk flows due to how the coordinate directions point relative to the direction of the SDSS data. }
    \label{fig:comparison_sdss_mocks_nonlinear_zeldovich}
\end{figure}

We have also applied the Kaiser MLE to subsets of the CF4 mocks, by applying it individually to the individual SDSS, CF4TF and 6dFGSv mocks that make up the CF4 mocks, with the same cut--off scales applied to the datasets. The results can also be found \hyperlink{https://github.com/abbew25/Measuring\_bulkflows}{here}. A similar bias with the trend of underestimating the underlying bulk flow also appears to afflict these results. As the bias was not overtly present in the results for the Zeldovich mocks for SDSS or the more generic zeldovich mocks tested earlier, this might suggest that the presence of the non--linearities in the PV field (for realistic mocks) have some influence on the results after all. This is demonstrated here in Figure~\ref{fig:comparison_sdss_mocks_nonlinear_zeldovich}, which shows results when applying the Kaiser MLE to mocks for the SDSS data where non--linear information is included in the peculiar velocity field (top panels) vs when it is not because the Zeldovich approximation is used to generate the mocks only (lower panels). 

Interestingly, the bias appears to possibly be more significant in the mocks with the CF4TF or 6dFGSv data only, which includes measurements at lower redshifts, where we might expect the systematic error in each measurement of the log--distance ratio to be less signifcant (due to the way uncertainty increases with more distant objects), thus allowing errors in modelling to dominate. Furthermore, as SDSS has the largest amount of data more non--linear information can be smoothed out. While we have discussed that the non--linear PVs do not appear to correlate with the $\chi^2$ goodness--of--fit from applying the Kaiser MLE, they are still potentially a source of systematic bias, because the Kaiser MLE method does not account for non--linearities in the PV field. Alternatively, the choice of $\sigma_* = 300 \mathrm{km s}^{-1}$ requires variation. This could be tested by applying a MCMC approach or the $\eta$MLE approach shown in \cite{qin2018bulk}, which might even have the affect of improving the $\chi^2$, but we leave this to future work to determine. 

Furthermore, the uncertainty due to the data for each estimate appears significantly underestimated as we obtain a reduced $\chi^2 \sim 23$ in general. We also find if a correction is applied to the bulk flows to correct the bias, the reduced $\chi^2$ is $\sim 24.8$ (where we simply correct $\mathbf{\mu} = \mathbf{x} - \mathbf{x_t}$, where $\mathbf{x}$ represents the vector of bulk flows compared to the truth bulk flows $\mathbf{x_t}$) in the equation for $\chi^2 = \mu \mathbf{C}^{-1} \mu^T$, where $\mathbf{C}$ is the covariance matrix. Therefore we will follow the approach of \cite{howlett2022sloan} and apply a scaling to the covariance matrix $\mathbf{C}$ for $\mathbf{\tilde{B}_{\mathrm{K}}}$ for the uncertainty on the real CF4 data when applying the Kaiser MLE method.  

\subsubsection{Peery MVE method: results for CF4 mocks}

\begin{table}
    \centering
    \caption{Results for the reduced $\chi^2$ goodness--of--fit for the bulk flow when applying the Peery MVE for 512 mocks for the CF4 data, compared to the chosen radius $R$ of the ideal survey used for the method.}
    \begin{tabular}{p{1.45cm}|p{0.25cm}| p{0.25cm} | p{0.25cm} | p{0.25cm} | p{0.25cm} | p{0.25cm} | p{0.25cm} | p{0.25cm} | p{0.25cm}} \hline 
       $R$ Mpc $h^{-1}$ & 69 & 104 & 138 & 173 & 207 & 242 & 276 & 311 & 345 \\ \hline 
       $\chi^2$ & 6.51 & 5.07 & 4.23 & 3.78 & 3.54 & 3.43 & 3.39 & 3.39 & 3.39 \\ \hline 
    \end{tabular}
    \label{tab:results_peerychisquare_mocks}
\end{table}

We also apply the Peery MVE method to 512 mocks. We also wanted to test the results on the mocks when the radius of the ideal survey for this method was varied between scales of 35 and 345 Mpc $h^{-1}$ (for clarity here, we do not apply any cuts to the data but only change the ideal survey geometry, unlike the cutoff scales we introduce when testing the Kaiser method on these mocks). Likewise, this is to look at the performance on mocks before applying the same procedures to the real data. This is also what was done in \cite{watkins2023analyzing} for the CF4 dataset, but with different choices for the ideal survey radii. In general we found the results were strongly correlated to the expected bulk flow moments of the mocks, regardless of the choice of the radius of the ideal survey, although the correlation between the $B_{i,\mathrm{P}}$ and $\tilde{B}_{i,\mathrm{P}}$ is not as strong as in the application of the Kaiser method. Likewise, there is a similar bias in that the estimated bulk flow is an underestimate of the expected bulk flow moment. However the reduced $\chi^2$ for each set of mocks is much closer to unity compared to the Kaiser result and varies depending on the radius of the ideal survey. The reduced $\chi^2$ is found to obtain closer values to unity at larger radii for the ideal survey; the relationship between these is summarised in Table~\ref{tab:results_peerychisquare_mocks}. In the case the biases present in the mocks are corrected, we find for the case the ideal survey radius is set to $173 \mathrm{Mpc} h^{-1}$, we obtain $\chi^2 \sim 4.48$. Results of applying the Peery MVE to 512 mocks with the ideal survey radius set to 173 Mpc$h^{-1}$ are shown in Figure~\ref{fig:results_peery_realisticmocks_CF4}. The results here (and in the Kaiser approach) will be used to scale the error bars appropriately when applying this method to the real data, which is described at the beginning of section~\ref{results_CF4data_thiswork}.

\subsubsection{Effects of systematic zero--point offsets}

Before finally applying estimators to the real data, we explore the effects of systematic offsets in the zero--point of datasets on the recovered bulk flows from the Kaiser method and the Peery MVE method. We explore both a global zero--point offset, in which case each log--distance ratio in the dataset is offset by a constant amount, $\sigma_{\eta}$ and a zero--point offset between the three main subsamples of CF4; SDSS, 6dFGSv and CF4TF. A global offset in the zero--point is analagous to a change or error in the Hubble constant $H_0$. It may also be possible to have an offset due to a calibration error when fixing the zero--points of datasets that have been combined.

Given the constraint equation that is included by \cite{peery2018easily} in order to ensure the results of the Peery MVE are independent of the Hubble constant, we can expect that this method should obtain a bulk flow measurement that is unaffected by a global zero--point offset in the data. We found that by applying the Peery MVE to the mocks with a deliberately added or subtracted global zero--point offset of $\sigma_{\eta} = 0.031$ (corresponding to an change of approximately 5 km $\mathrm{s}^{-1}$ $\mathrm{Mpc}^{-1}$ on $H_0$) the estimated bulk flow remains unbiased, validating this expectation. 

\begin{figure*}
    \centering
    \includegraphics[width=1.0\textwidth]{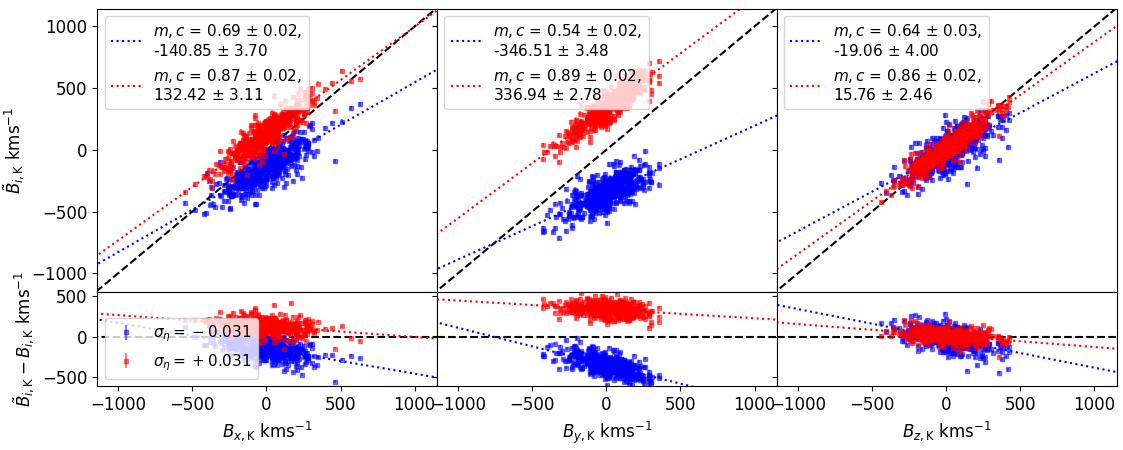}
    \caption{The recovered bulk flow moments $\tilde{B}_{i,\mathrm{K}}$ and the residuals $\tilde{B}_{i,\mathrm{K}} - B_{i,\mathrm{K}}$ in Supergalactic coordinates, compared to the expected truth bulk flow moments $B_{i,\mathrm{K}}$ for the CF4 mocks, when applying the Kaiser MLE. The dashed line shows a 1--to--1 line in the plots showing $\tilde{B}_{i,\mathrm{K}}$ vs $B_{i,\mathrm{K}}$, and a horizontal line for the residual plots to compare to the results. The red dashed lines show best fit linear regressions for $\tilde{B}_{i,\mathrm{K}}$ vs $B_{i,\mathrm{K}}$ in each panel. The legend indicates the gradient $m$ and the y--intercept $c$ for each regression. 
    The blue points show the results when there is a global zero--point shift of $\sigma_{\eta} = -0.031$ applied to each object log--distance ratio in the mocks. For the red points there is a global zero--point shift of $\sigma_{\eta} = +0.031$ applied to each object. The change in the linear regressions in columns of panels allow us to measure the affect of the change in the zero--point on the results.}
    \label{fig:kaiser_ZPglobalmocks_results}
\end{figure*}

However, we have no such expectation when applying the Kaiser MLE estimator and find that the results that are presented (in Supergalactic coordinates, to match the coordinates we present the data results in) show significant biases when a global zero--point offset is present; this is shown in Figure~\ref{fig:kaiser_ZPglobalmocks_results}. In particular the Supergalactic $y$ direction, which is the direction of the SDSS cone of data, is affected the most significantly. The Supergalactic $x$ and $z$ directions are also visibly affected due a lesser extent. Due to the fact a constant shift in each $\eta$ does not translate to a constant shift in each PV, the effect of a positive or negative $\sigma_{\eta}$ is not symmetric. Overall the Bulk Flow amplitude appears to be shifted for the $x$ and $y$ directions by upward of 100 $\mathrm{km s}^{-1}$.

\begin{figure*}
    \centering
    \includegraphics[width=1.0\textwidth]{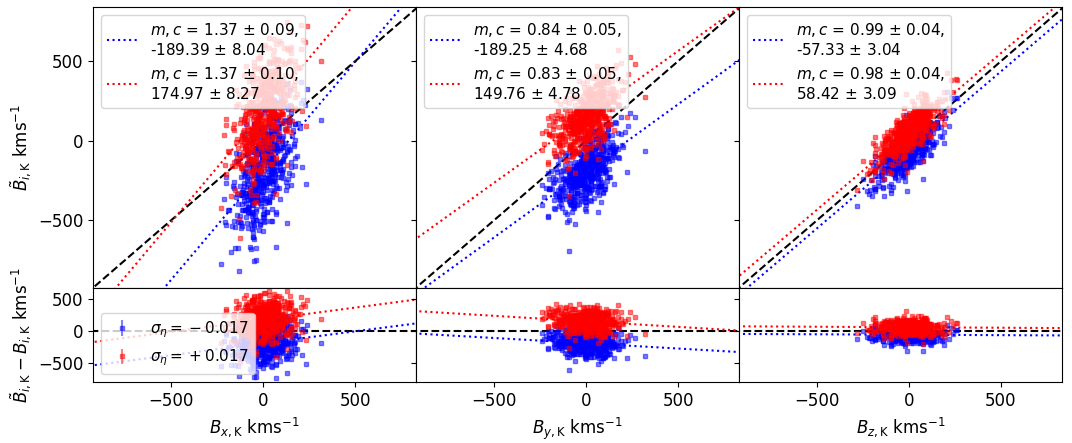}
    \caption{The recovered bulk flow components $\tilde{B}_{i,\mathrm{P}}$ and the residuals $\tilde{B}_{i,\mathrm{P}} - B_{i,\mathrm{P}}$ in Supergalactic coordinates, compared to the expected truth bulk flow moments $B_{i,\mathrm{P}}$ for the CF4 mocks, when applying the Peery MVE. The dashed lines shown are the same as described in Figure~\ref{fig:kaiser_ZPglobalmocks_results}. For the blue points there is a relative zero--point shift of $\sigma_{\eta} = -0.017$ applied to CF4TF object log--distance ratios in the mocks only. For the red points there is a relative zero--point shift is $\sigma_{\eta} = +0.017$.}
    \label{fig:peery_ZPmocks_results_cf4tfrelshift}
\end{figure*}

We also explore the effect of a relative zero--point offset between the different datasets that make the CF4 mocks; because the CF4 mocks are composed of individual mocks made from SDSS, CF4TF and 6dFGSv this is straight forward to implement. We choose a systematic offset of $\sigma_{\eta} = \pm 0.017$ between different datasets; this value is chosen based on the statistical uncertainty found in the calibration procedure of different datasets in \cite{tully2023cosmicflows}. In Figure 16 of \cite{tully2023cosmicflows} work, the 1--$\sigma$ statistical uncertainty on the relative zero--point offset between the FP and TF datasets with the SNe Ia used to calibrate the entire CF4 dataset is five times smaller than this chosen value; we choose a larger value in order to simulate the effect of a significant (5$\sigma$) systematic error. We first apply this offset to the CF4TF mock data relative to all other data and then to the 6dFGSv mock data relative to all other data. We find that when applying these offsets there is a visible bias for the recovered bulk flows (as expected) when applying both the Kaiser MLE and the Peery MVE. Interestingly the Peery MVE seems to be more greatly affected by a relative offset between datasets than the Kaiser MLE estimator. Figure~\ref{fig:peery_ZPmocks_results_cf4tfrelshift} shows the effect of a relative zero--point offset added to the CF4TF data in the CF4 dataset and how this affects the results from the Peery MVE. More of these results can be found \hyperlink{https://github.com/abbew25/Measuring\_bulkflows}{here}. Of all individual offsets we test with different estimators, the largest bias is apparent in the Supergalactic $y$ direction when a global offset is added to the CF4 mocks and the Kaiser MLE approach is used to recover the bulk flow, shown in Figure~\ref{fig:kaiser_ZPglobalmocks_results}.

\subsection{Summary: mock performance evaluation}

To summarise this section, we found the following;

\begin{itemize}
    \item The Watkins MVE method and Kaiser MLE are able to recover the expected bulk flow moment of a dataset accurately and with good precision, when the bulk flow is a constant valued vector across the survey volume.
    \item In general however, these estimators may not measure the bulk flow moment with the same accuracy and precision for realistic mocks, and the measured moment is not in general consistent with Equation~\ref{eq::definition_bulkfflow}. This issue was first pointed out by \cite{nusser2014inconsistency}. We demonstrate the precision of the estimators relative to the expected bulk flow moments thoroughly here with mocks. 
    \item We show there is a strong dependence between the survey geometry (in particular the selection function) and the $\chi^2$ goodness--of--fit of these methods.
    \item Both \cite{nusser2014inconsistency} and the \cite{peery2018easily} introduce modifications into the Kaiser MLE method and Watkins MVE method to address this issue. The Nusser MLE and the Peery MVE method appear to give a better fit to the expected bulk flow moment, that is consistent with Equation~\ref{eq::definition_bulkfflow}, with a $\chi^2$ closer to unity for different survey geometries we have tested and also reduce the correlation between survey geometry and the $\chi^2$. However the Nusser MLE modification depends on the assumption of spherical symmetry, which we found to be unsuitable for the CF4 dataset.
    \item For mocks with highly non--linear PVs the PV distribution changes and this appears to introduce a bias to the recovered bulk flows from any estimator we test. This may be more apparent in the case the systematic errors in the observed PV are smaller, there are less PV observations, and error due to modelling only linear--theory velocities begins to dominate the measurement. Therefore a more complete solution may need to be derived in future work to be able to estimate bulk flows from data that are both comparable to theory and which generally obtain an accurate estimate independent of the survey geometry or the peculiar velocity distribution. We leave this task to future work. 
\end{itemize}

When using the estimators we have tested in this work, we suggest they \emph{always} need to be applied to realistic mocks before applying them to estimate the bulk flow of a real dataset and compare the results to theory. This is because our work has shown one needs to quantify how well the calculated error for the bulk flow due to the data is actually capturing the uncertainty that should be assigned to the bulk flow, and test for the presences of biases. If a bias is present, a correction to the bulk flow can be made before comparing the measurement to theory. For this reason we have applied the Kaiser MLE and Peery MVE to realistic mocks for the CF4 dataset, which will allow us to draw a more accurate conclusion about 1) the uncertainty on the measured bulk flow vector using these methods and 2) the level of tension or agreement the bulk flow has with theoretical expectations. Based on our results, a correction will be applied to the measured bulk flows from these estimators using the CF4 data and the uncertainties on the measurement will be scaled to recover a reduced $\chi^{2}$ of 1. 

For measurements applied to the real data, one may be concerned about the effects of selection bias or Malmquist bias which are well described in \cite{strauss1995density}. As mentioned previously, galaxies are placed at their observed redshifts in order to determine their distances in all the mocks we test and the same approach is taken to get galaxy distances for the real data. Using the galaxy redshifts as opposed to using a distance indicator to determine the distances may cause the measured bulk flows to suffer from selection bias, and less so from Malmquist bias. However, since our realistic mocks for CosmicFlows--4 include both the selection function and clustering for the real data, conclusions we draw from applying the estimators to the mocks should also be valid for the real data.

\section{Measured bulk flows from real data}\label{sec::results}

In this next section we apply the Kaiser MLE scheme and the Peery MVE scheme to the CF4 dataset and compare the measured bulk flows to the expectation of the $\Lambda$CDM model, which we deduce from resulting bulk flows of mocks for the CF4 data where we have applied these schemes using the true radial PVs of each object in the mock with zero--uncertainty. We note here an analysis of the CF4 dataset using the Peery MVE appeared in preprint in \cite{watkins2023analyzing} while we were completing this analysis. However we still present our results here for both the estimators as we expect our conclusions regarding tension with the $\Lambda$CDM model will differ. While we expect to obtain a similar bulk flow amplitude, we argue the error bars presented for the result in \cite{watkins2023analyzing} are underestimated given our results from applying the estimator to mocks, and that their corresponding claim of a tension with $\Lambda$CDM is slightly overstated.

\subsection{Results: CosmicFlows 4 dataset}\label{results_CF4data_thiswork}

\begin{table*}
    \centering
    \caption{Estimated bulk flow components in Supergalactic coordinates from the Peery MVE and the Kaiser MLE. $d_e$ gives the effective depth of each bulk flow estimate, and $P(> \chi^2)$ gives the probability of obtaining a larger $\chi^2$ goodness--of--fit for the measured bulk flow with respect to the expectation from the $\Lambda$CDM model. $\langle|\mathbf{B}|\rangle$ gives the mean bulk flow estimated from 512 CF4 mocks when the true radial peculiar velocities with zero uncertainty have been used for each object in the mock, as a proxy for comparison to the expectation to $\Lambda$CDM, and the error bar represents the standard deviation of this measurement. The error bars include both error due to each object's PV uncertainty and cosmic variance. The columns for results that specify `corrected' indicate the bulk flow modes and amplitude have been corrected for bias based on the simulations for the estimators. For all columns, the numbers in brackets for error bars and probabilities indicate the result if no scaling is applied to the systematic uncertainty based on the $\chi^2$ from the mocks. }
    \begin{tabular}{c|c|c|c|c} \hline 
         & Kaiser MLE & Peery MVE & Kaiser MLE (corrected) &  Peery MVE (corrected) \\ \hline 
        $\tilde{B}_x$ $\mathrm{km s}^{-1}$& -281 $\pm$ 164 (153) & -328 $\pm$ 100 (83) & -382 $\pm$ 165 (153) & -391 $\pm$ 104 (83) \\
        $\tilde{B}_y$ $\mathrm{km s}^{-1}$& 37 $\pm$ 148 (142) & -102 $\pm$ 89 (75) & 48 $\pm$ 149 (142) & -119 $\pm$ 93 (75) \\
        $\tilde{B}_z$ $\mathrm{km s}^{-1}$& -105 $\pm$ 152 (140) & -94 $\pm$ 119 (108) & -135 $\pm$ 154 (140) & -126 $\pm$ 122 (108) \\
        $|\mathbf{\tilde{B}}|$ $\mathrm{km s}^{-1}$ & 302 $\pm$ 164 (153) & 357 $\pm$ 104 (87) & 408 $\pm$ 165 (153) & 428 $\pm$ 108 (87) \\
        $d_e$ $\mathrm{Mpc}$ ${h}^{-1}$& 49 & 173 & 49 & 173 \\
        $P(> \chi^2)$ & 32.7\% (26.6\%) & 0.71\% (0.06\%) & 10.1\% (6.59\%) & 0.11\% (0.002\%) \\ \hline 
        $\langle|\mathbf{B}|\rangle$ $\mathrm{km s}^{-1}$ ($\Lambda$CDM) & 130 $\pm$ 54 & 128 $\pm$ 59 & 196 $\pm$ 82 & 194 $\pm$ 86\\ \hline 
    \end{tabular}
    \label{tab:results:bulkflows_data}
\end{table*}

\begin{figure*}
    \centering
\includegraphics[width=1.0\textwidth]{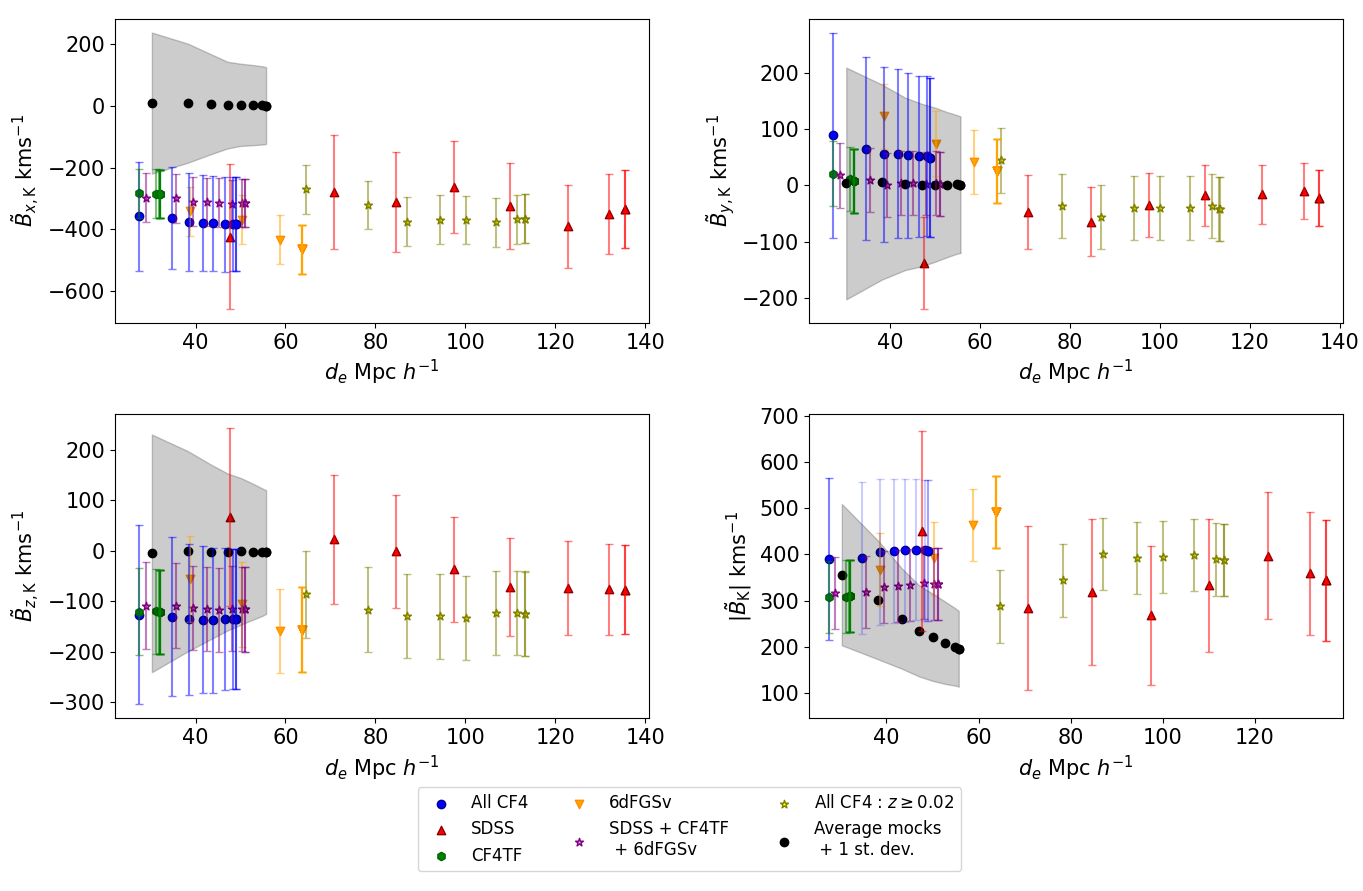}
    \caption{Estimated bulk flow components in Supergalactic coordinates and bulk flow amplitude from the CF4 data and subsets of it, using the Kaiser MLE. The bulk flow has been estimated at various different effective depths (the same effective depths are plotted for each panel) by applying cuts to the data at different radii and calculating the effective depth. The error bars do not include cosmic variance for clarity. The different data subsets are shown in the legend. The black points with grey shading show the average recovered bulk flow from 512 mocks $\langle |\mathbf{B}_{\mathrm{K}|} \rangle$ with the same selection function as SDSS+CF4TF+6dFGSv, and with the true radial PVs of the objects fed as input. The shaded region shows the standard deviation of $|\mathbf{B}_{\mathrm{K}|}$ which gives an estimate of the theoretical cosmic variance uncertainty for the mocks. }
    \label{fig:kaiser_results_data}
\end{figure*}

\begin{figure*}
    \centering
    \includegraphics[width=1.0\textwidth]{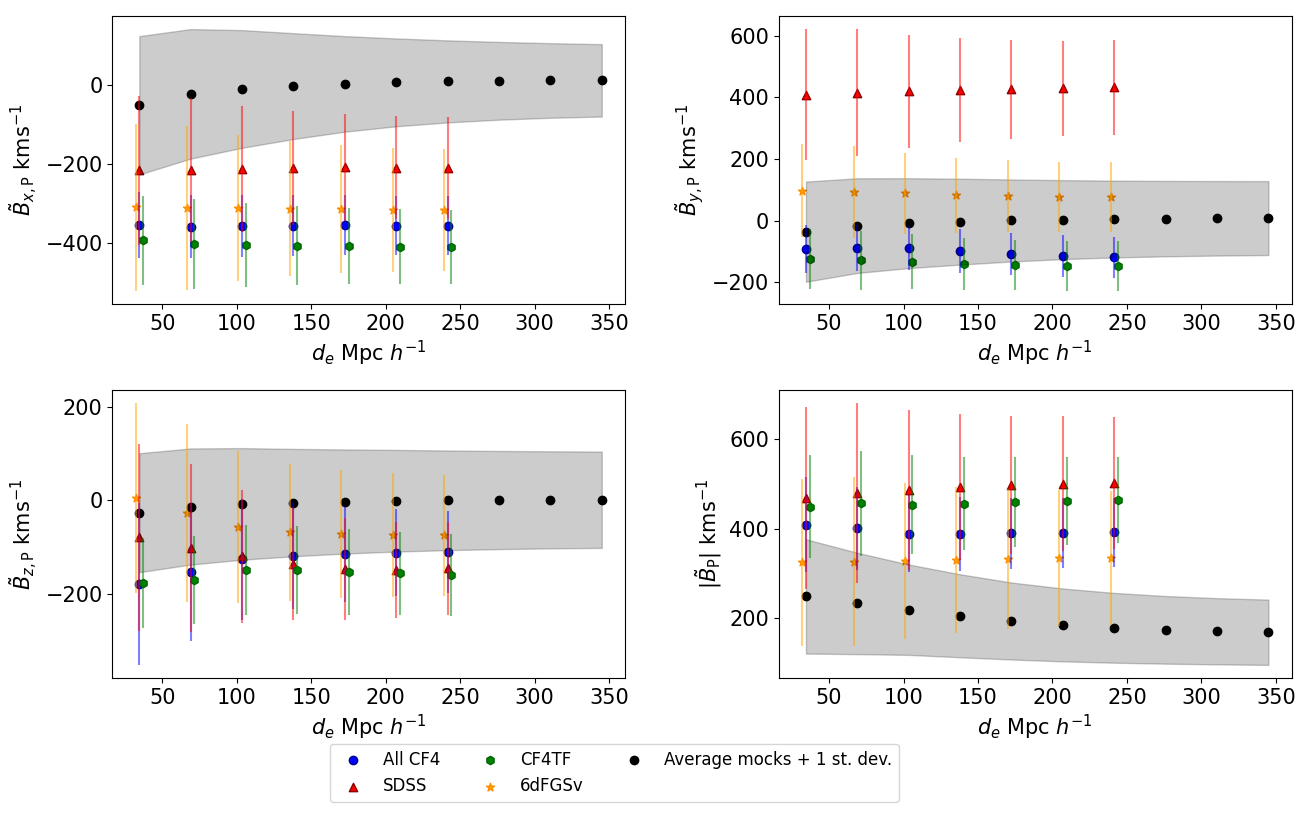}
    \caption{As in Figure~\ref{fig:kaiser_results_data}, but for the estimated bulk flow components from the Peery MVE and where the different effective depths for estimates arise by changing the radius of the ideal survey for this estimator instead of cutting the data at different scales. A small horizontal offset is added to the CF4TF data points and 6dFGSv data points about their correct value for $d_e$ so that datapoints do not overlap. }
    \label{fig:mve_results_data}
\end{figure*}

\begin{figure*}
    \centering
    \includegraphics[width=1.0\textwidth]{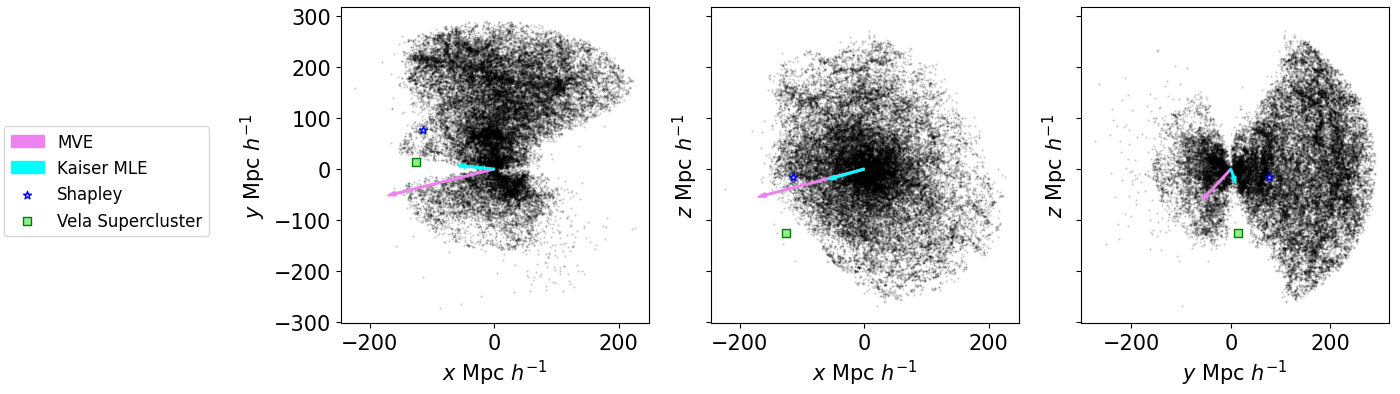}
    \caption{A map of the CF4 dataset in Supergalactic coordinates, each panel showing a 2D projection of two coordinates. The arrows positioned at the origin show the direction of the bulk flow (with relative lengths given by the amplitude of the bulk flow in each direction, scaled by the effective depth of the measurement) for the Peery MVE and Kaiser MLE. We also show the approximate positions of Shapley \protect\citep{proust2006shapley} and the hypothesised Vela Supercluster \protect\citep{kraan2017discovery} . }
    \label{fig:3d_bulkflow_plot_SGcoords}
\end{figure*}

Table~\ref{tab:results:bulkflows_data} shows the estimated bulk flow components for the CF4 dataset when applying the Peery MVE and the Kaiser MLE, in Supergalactic coordinates. The error bars for each component due to the uncertainty in the data have been scaled to account for how well the bulk flow estimators perform on mocks. This is done so that the reduced $\chi^2$ will rescale to unity, as $C_{ij} \rightarrow \beta C_{ij}$ where $\beta$ is the rescaling factor set by the reduced $\chi^2$ from the mocks, in order to inflate the error on the covariance matrix of the PV uncertainties. $C_{ij}$ here refers to the covariance matrix component due to uncertainty on the PVs only and not due to cosmic variance uncertainty. Uncertainties in brackets are given to indicate what the uncertainty is without any scaling applied. 

It should be noted that the uncertainty due to cosmic variance (CV) may be imprecise because it is calculated based on linear theory, yet we can understand that bias may arise in the results due to a lack of modelling for non--linear structure growth. On may consider to instead estimate the CV covariance of the bulk flows from the ($\Lambda$CDM) mocks. However, we expect this to slightly underestimate the true CV uncertainty for the Kaiser method because the mocks do not include a small portion of lower redshift objects such as Ia SNe, SBFs, etc. that are included in the CF4 data,\footnote{As a reminder, it was mentioned previously that the mocks we use for the CosmicFlows--4 dataset are created by stacking mocks that were produced for the SDSS data, CF4TF data and 6dFGSv data that largely capture the survey geometry and data included in the CosmicFlows--4 catalogue.} and thus will obtain a CV uncertainty corresponding to a slightly lower effective depth. However we find that the linear theory CV covariance matrix and the CV covariance of the mocks only differs by a factor of $\sim 1.2$ for the Kaiser method. Likewise for the Peery MVE, we find that the linear theory CV covariance matrix is also similar to the CV covariance matrix from the mocks, only differing by a factor of $\sim 0.7$. Therefore we simply use the linear theory results to calculate the total error on the measured bulk flows from the data and for comparison to the $\Lambda$CDM model. Nonetheless, the mocks form a useful basis for comparison, so in table~\ref{tab:results:bulkflows_data} we also include $\langle|\mathbf{B}|\rangle$, the mean bulk flow estimated from 512 CF4 mocks when the true radial peculiar velocities with zero uncertainty have been used for each object in the mock. 

To compare our measured bulk flows to the $\Lambda$CDM model, we take our linear theory covariance and use it to compute the $\chi^2$ of our measurements and have listed the probability of finding a larger $\chi^2$ in Table~\ref{tab:results:bulkflows_data}. For the effective depths in the table, we use the mean distance to each object weighted by $(\sigma_n^2 + \sigma_*^2)^{-1}$ for the Kaiser MLE, while for the Peery MVE we use the radius of the ideal survey for the estimator, which is set to $\sim 173 \mathrm{Mpc} h^{-1}$ in our results. Furthermore, we have added to the table the expected bulk flow components and amplitude after the bias in the mocks has been corrected in additional columns. We also apply the relevant correction to the $\Lambda$CDM expectation calculated from the mocks, $\langle|\mathbf{B}|\rangle$, for comparison. 

It is interesting that both estimators give significant bulk flow magnitudes in the negative Supergalactic $x$ direction. While the amplitudes of each measurement from the different approaches are in agreement within statistical uncertainty the Kaiser result would indicate there is no tension with the $\Lambda$CDM model. In contrast, the result from the Peery MVE, which is able to probe the bulk flow at a larger effective depth indicates there is tension with $\Lambda$CDM if we set the significance level at 5\%. While we find a similar magnitude for the bulk flow to \cite{watkins2023analyzing} the tension we find with the $\Lambda$CDM prediction is slightly less significant after correcting the error bars and the bias in the results; we find the bulk flow has a 0.11\% chance of occurring rather than only the 0.0002\% or less they determine.

To try and investigate the origin of this, Figure~\ref{fig:kaiser_results_data} shows the result of applying the Kaiser MLE to the CF4 mocks (and subsets of the CF4 data with the same zero--point calibration) with the same radial cuts as was applied to the mocks. No bias correction has been applied in the results shown, but the values for each CF4 result (blue points) can still be fairly compared to the expected values given from the mocks without bias correction. 

One may note the effective depth for the mocks with the same radial cuts (shown in every panel on the $x$--axis) are generally larger than for the real data; this can be attributed to the fact the mocks do not include data for SNe Ia or other low redshift objects used to calibrate the dataset, as mentioned previously. As these objects typically have PV measurements with smaller errors they carry larger weight for the Kaiser MLE approach. The lower redshift CF4TF data also influences the effective depth of the CF4 data enough for it to sit at lower depths than is obtained for the SDSS data or 6dFGSv data alone; this is seen by the comparing results for all CF4 data to the subset made up of 6dFGSv + SDSS + CF4TF only (purple stars) and then the SDSS subsets (red triangles) and 6dFGSv subsets (orange triangles). The subset of CF4 with only $z \geq 0.02$ objects included has a significantly larger effective depth as much of the contribution of CF4TF catalogue is removed. 

Almost all of the data in Figure~\ref{fig:kaiser_results_data} shows unexpected trends for the bulk flow magnitude as a function of depth given the expectation from $\Lambda$CDM, which should be similar to what we see in the mock results. Instead of the bulk flow amplitude decreasing with depth, it increases, particularly in the case of the 6dFGSv subset. The clearest feature is that all subsets of data we test prefer a large negative bulk flow in the Supergalactic $x$-direction at all effective depths, which is the primary driver of the large bulk flow amplitude.

Figure~\ref{fig:mve_results_data} shows a similar plot to Figure~\ref{fig:kaiser_results_data} however this time with the Peery MVE. For the CF4 data, the bulk flow amplitude has been evaluated at significantly different scales to the Kaiser MLE but the resulting components and amplitude are not significantly different. Interestingly however, this is with the exception that the Peery MVE obtains a much larger bulk flow amplitude in the positive Supergalactic $y$ direction for the SDSS data, although the direction is negative for the entire CF4 data. We may expect that is related to the fact that SDSS has most of its data in the positive Supergalactic $y$ direction of the sky in a cone--like geometry. In contrast, the Peery MVE seems to give a more reasonable estimate for the bulk flow of the 6dFGSv data alone, compared to the larger bulk flow amplitude of 6dFGSv estimated to be close to $\sim 500 \mathrm{km s}^{-1}$ from the Kaiser MLE at the largest scales. It is apparent that at larger depths for the CF4 dataset, at which we expect the estimated bulk flow to be more accurate given the mock results for the Peery MVE, the resulting bulk flows seems to diverge from the expectation of the $\Lambda$CDM model (the black points and grey shaded regions). Although, individual datasets present differences in their bulk flows at different depths between the Kaiser MLE and Peery MVE estimators, the overall bulk flow amplitudes are remarkably consistent in their deviation from $\Lambda$CDM.

Figure~\ref{fig:3d_bulkflow_plot_SGcoords} shows a map of the CF4 data and the direction of the estimated bulk flows given in Table~\ref{tab:results:bulkflows_data} (a bias correction is applied in the data shown). The results presented in this work would indicate that in general the bulk flow is headed towards the hypothesised Vela Supercluster \citep{kraan2017discovery} close to the general region of the `Great Attractor' \citep{radburn2006structures}, and with some pull in the direction of the Shapley supercluster. The former of these hides out of sight in the Zone of Avoidance, where the Milky Way disk inconveniently blocks the view of extragalactic objects, while the latter remains only partially covered by current data in the southern hemisphere.

The strange trends we see in the data could arguably be a symptom of a systematic in zero--point offsets. However we have noted the bulk flow amplitudes and values of the of the bulk flow moments measured from the Kaiser MLE and the Peery MVE method are similar. We might expect if a global zero--point offset was present in the data and no relative zero--point offsets were present, the Kaiser MLE and Peery MVE would obtain differing amplitudes because we know the Peery MVE is effectively immune to this systematic via the additional constraint equation it incorporates to minimize uncertainty due to $H_0$. However, this would not rule out the possibility of errors due to an incorrect relative zero--point offset between the datasets that make up the CF4 data because they introduce a bias into the recovered bulk flow from each method in different ways. We might expect if a relative zero--point offset error affected the data in such a way that the recovered bulk flow mode value in the negative Supergalactic $x$ direction was overestimated, accounting for this error would make the bulk flow measurements more consistent with the expectation from $\Lambda$CDM. A zero--point offset in the CF4TF data of $\sigma_{\eta} = -0.017$ relative to SDSS and 6dFGSv appears to introduce a bias of $\sim -189 \mathrm{km s}^{-1}$ in the Supergalactic $x$ direction for the Peery MVE, which would work in the correct direction to account for the bulk flow in the data, and would also reduce the bulk flow amplitude by roughly the same amount in the SuperGalactic $y$ direction and by a smaller amount in the $z$ direction, and thus would reduce the tension seen in our results with $\Lambda$CDM. The same offset would also mean that the bulk flow measured by the Kaiser MLE would require a correction such that the bulk flow amplitude would become lesser in the $x$ and $z$ directions and slightly larger in the $y$ direction, although the required corrections are considerably smaller. However, this scenario does require a zero-point offset between datasets at a level of $5\sigma$ compared to the reported calibration uncertainty in CF4. Overall, we conclude that while we cannot rule out internal zero-point offsets systematically enhancing the bulk flow amplitude, these would have to be very large, unnoticed, and applied in a particularly contrived way to affect both estimators in the right directions by the right amount. 

Alternatively, we could also consider the findings in \citet{heinesen2023bulk} which shows that by neglecting the importance of relativistic effects in bulk flow measurements, the bulk flow amplitude may be overestimated by a factor of $\sim (1 + 1.55z)$, where $z$ here represents the redshift of the effective depth of the bulk flow measurement. This may also explain the trend of increasing bulk flow amplitude with effective depth which we see in our results, and would lead to an overestimate of approximately $\sim 10$\% for the Peery MVE result and $\sim 2.5$\% for the Kaiser MLE result. However, even after applying this correction, the Peery MVE result is still $389 $ km/s, which is still in tension with $\Lambda$CDM with only a $\sim 0.4$\% chance of measuring the observed bulk flow.

Finally, one may consider that there is some possibility the tensions seen in bulk flow measurements may be related to the $H_0$ tension \citep[see references within][]{di2021realm} and \citep[][]{aghanim2020planck, heymans2021kids} $\sigma_8$ tension that currently pervade cosmology. The theoretical bulk flow variance is defined as \citep{andersen2016cosmology}
\begin{equation}
    \sigma_{V}^2(\mathbf{r}) = \int \frac{dk^3}{(2\pi)^3} P_{vv}(k) | \Tilde{W}(\mathbf{k;r})|^2.
\end{equation}
which are sensitive to $H_0$ and $\sigma_8$. $P_{vv}(k)$ is the velocity power spectrum while $\Tilde{W}(\mathbf{k;r})$ is the Fourier space window function of the galaxy survey. The most probable bulk flow amplitude can be estimated as $V_p(R) = \sqrt{3/2} \sigma_V(R)$, and thus has dependence on $\sigma_8$ and $H_0$ through the power spectrum. However, altering the power spectrum to vary $H_0$ and $\sigma_8$ within the range seen by tensions only changes the resulting bulk flows in the $\Lambda$CDM model by small numerical factors; varying these parameters is not able to solve the tensions seen in bulk flow measurements. Furthermore, we would still expect even after allowing these parameters to vary that the bulk flow amplitude should approach zero at increasingly larger scales.  

\subsection{Comparison to bulk flows from previous literature}

\begin{figure*}
    \centering
\includegraphics[width=1.0\textwidth]{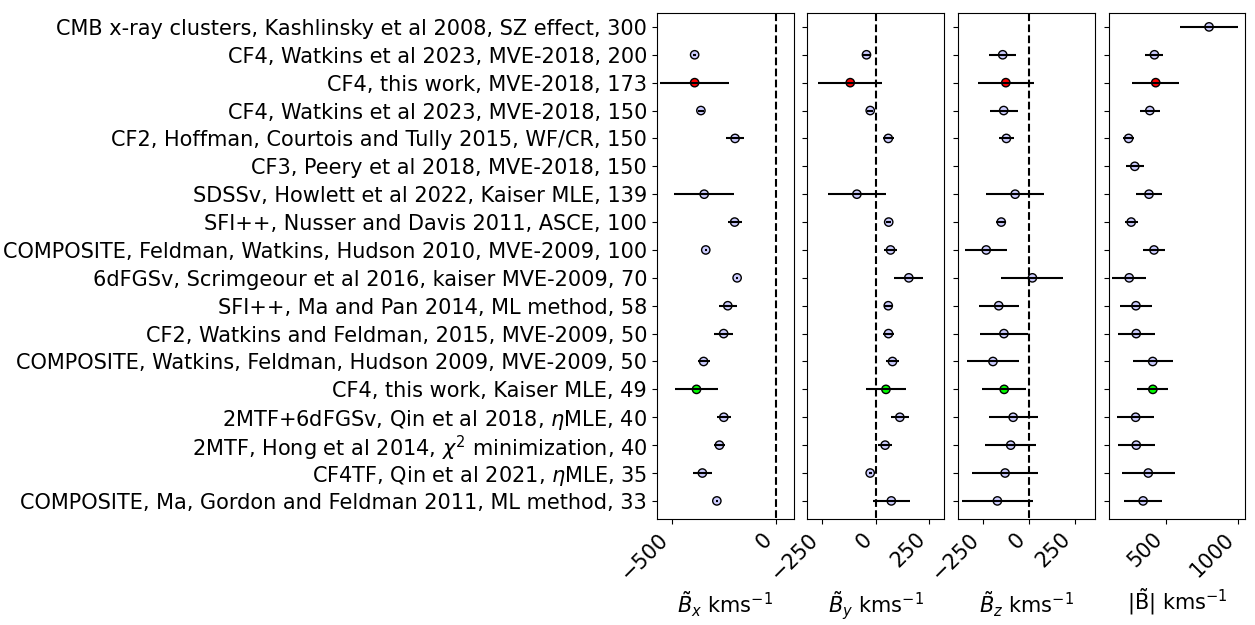}
    \caption{Comparison of different bulk flows of datasets shown in the literature, converted to Supergalactic coordinates and listed in order of decreasing effective depth. In cases where data is missing coordinates or bulk flow components were not explicitly listed in writing, only the bulk flow amplitude is shown. Error bars given for $|\tilde{B}|$ are exact, except for when the uncertainty due to cosmic variance was not given and we have included a rough estimate from theory for a perfect sphere with radius corresponding to the effective depth given. To obtain error bars for each Supergalactic coordinate mode we have assumed the correlation between modes in other coordinate systems was zero, as in general the covariance matrix was not available. For asymmetric error bars we took the mean of the two uncertainties to perform the transformation of the uncertainties to different coordinates. It should be noted that in general authors do not always show uncertainties in a consistent way depending on whether they used mocks to account for underestimated uncertainties by estimators or accounted for the survey geometry when computing the theoretical cosmic variance uncertainty. On the left hand side of the plot we list the dataset used, the authors, the method employed to obtain the measurement (MVE--2018 is short for Peery MVE, MVE--2009 for Watkins MVE, SZ effect for Sunyaev--Zeldovich effect, ML for maximum likelihood estimation, WF/CR for Wiener Filter/Constrained realizations, ACSE for the `all space constrained realizations' method ), then $d_e$ in Mpc $h^{-1}$. }
    \label{fig:bulkflows_comparison}
\end{figure*}

Finally we show a comparison of our bulk flow estimates to some of those previously found in the literature. This is shown in Figure~\ref{fig:bulkflows_comparison}, which includes results taken from \cite{ma2014estimation, hong20142mtf, qin2018bulk, qin2021cosmic, watkins2009consistently, feldman2010cosmic, howlett2022sloan, watkins2023analyzing, peery2018easily, scrimgeour20166df, watkins2015large, kashlinsky2008measurement, ma2011peculiar, nusser2011cosmological, hoffman2015cosmic}. The results are shown in order of increasing depth. It is interesting to note that there is a consistently large bulk flow detected in the negative Supergalactic $x$ and $z$ directions, despite the fact different data sets, methodologies and approaches to calibrate the zero--point have been used. However, one should consider that the results between datasets will also be correlated because they all contain some overlap in the data included. Interestingly, the effective depth of each measured bulk flow varies but the bulk flow appears to be consistent in the amplitude and direction. As stated previously, it is interesting to note that bulk flow measurements reported at larger effective depths are those that tend to find more tension with $\Lambda$CDM and our measurements of the CF4 bulk flow are consistent with this trend. Although not included in Figure~\ref{fig:bulkflows_comparison}, \cite{migkas2021cosmological} also finds an apparent spatial variation in $H_0$ from galaxy cluster scaling relations which could be interpreted as a bulk flow of amplitude $\sim 900 \mathrm{km s}^{-1}$ at depths of $\sim 500$ Mpc which would also coincide with similar directions to the other bulk flows in the literature. 

The results in this work would indicate and support previous results in the literature that galaxies are headed towards the Greater Attractor or Vela Supercluster region in the negative Supergalactic $x$ direction. This strongly indicates that future surveys aimed towards the negative Supergalactic $x$ direction (primarily the southern equatorial hemisphere) are needed in order to gain a better understanding of bulk flow measurements and their tension with $\Lambda$CDM.

\section{Conclusions}\label{sec::conclusions}

In this work we have discussed the potential pitfalls on bulk flow estimators, and thoroughly evaluated how well the Kaiser maximum likelihood estimator \citep[Kaiser MLE;][]{kaiser1988theoretical}, Nusser MLE \citep{nusser2014inconsistency}, minimum variance estimator (MVE) by \cite{watkins2009consistently} and the Peery MVE \cite[][]{peery2018easily} are able to estimate the bulk flow, using mock data. This work has built on analysis with mock data already conducted for these estimators by others including \cite{agarwal2012testing, nusser2014inconsistency, andersen2016cosmology, qin2018bulk, qin2019redshift, qin2021cosmic, howlett2022sloan}. 

Overall, the main significant finding of our tests on mock data is that both the Kaiser MLE and Watkins MVE recover estimates of the bulk flow with a goodness--of--fit typically corresponding to a reduced $\chi^2 > 1$,  which is strongly correlated with survey geometry. This implies that these estimators are not able to capture the real systematic uncertainty in the bulk flow. We found that the modified weighting scheme to the Kaiser MLE scheme by \cite{nusser2014inconsistency} appears to perform better at estimating the bulk flow (the $\chi^2$ is closer to unity and the correlation between the survey geometry and $\chi^2$ appears to approach zero), but it is not suitable for surveys that do not have spherical symmetry in the selection function, which is the case for most realistic datasets. In contrast the Peery MVE also obtains a $\chi^2$ closer to unity in general and does not rely on the assumption of spherically symmetric survey geometry; arguably, this estimator is the most robust and reliable. However, all these estimators suffer from some bias due to highly non--linear PVs and some underestimation of the statistical error remains. Both of these need to be corrected (as is done herein) before comparing a bulk flow measurement to a cosmological model.

Our recommendations are that either 1) in future work one could develop an improved bulk flow estimator to overcome the issues identified here or 2) that when estimating the bulk flow, ensure the estimator is always tested on realistic mocks to quantify how well it performs given the survey geometry, to ensure the systematic error bars on the bulk flow estimate are accurate. We have taken the latter approach in regards to making an estimate of the bulk flow of the CosmicFlows--4 data using the Kaiser MLE \citep{kaiser1988theoretical} and the Peery MVE \citep{peery2018easily}.

We analysed CosmicFlows--4 mock data to determine the real systematic uncertainty in the bulk flow and thus scaled the uncertainties up according to the reduced $\chi^2$ fit from the mocks. Using the Kaiser MLE we found a bulk flow with an amplitude of 408 $\pm$ 165 $\mathrm{km s}^{-1}$, at an effect depth of $49$ $\mathrm{Mpc} h^{-1}$ that appears to not be in tension with the $\Lambda$CDM model. However our bulk flow estimate from the Peery MVE at a depth of $173$ $\mathrm{Mpc} h^{-1}$ is in tension with the model with an amplitude of 428 $\pm$ 108 $\mathrm{km s}^{-1}$, compared to the expected $\sim 192$ $\mathrm{km s}^{-1}$ based on $\Lambda$CDM mocks. The bulk flow amplitude is in excellent agreement with the measurement by \cite{watkins2023analyzing} who used the same method. However, our tension with $\Lambda$CDM is very slightly less severe because we derive larger error bars based on the mock results. 

Overall, our results indicate (along with previous consistent results in the literature) that the volume of galaxies in our local region is headed towards the Great Attractor \citep{radburn2006structures} or the Vela Supercluster \citep{kraan2017discovery} regions, with some influence from the Shapley supercluster. To gain a better understanding of our bulk flow and test the $\Lambda$CDM model more rigorously, it is necessary to gain more peculiar velocity data towards the negative Supergalactic $x$ direction (southern equatorial hemisphere) and further improve bulk flow estimators. 

\section*{Acknowledgements}

This research has made use of NASA's Astrophysics Data System bibliographic services, the astro-ph pre-print archive at \hyperlink{https://arxiv.org/}{https://arxiv.org/} and the python libraries \texttt{MATPLOTLIB}, \texttt{ASTROPY} \citep{hunter2007matplotlib, robitaille2013astropy}. We have also made use of the University of Queensland's Getafix, Tinaroo and Bunya high performance computers to conduct this research. The authors would also like to acknowledge Fei Qin for creating the Cosmicflows4 Tully--Fisher (CF4TF) mocks we used in this work, and AW would like to thank Anthony Carr for shared code used in this project. Both TD and CH acknowledge support from the Australian Government through the Australian Research Council's Laureate Fellowship funding scheme (Fl180100168). We also thank Eoin Ó Colgáin, Richard Watkins, Hume Feldman and the anonymous reviewer for their useful comments and thoughts on the arXiv manuscript.

\section*{Data Availability}

The CosmicFlows4 data used to estimate the bulk flow comes from the Extragalactic Distance Database (EDD) which can be found here: \hyperlink{https://edd.ifa.hawaii.edu/}{https://edd.ifa.hawaii.edu/}. The codes used to produce the bulk flow estimates in this work can also be found here: \hyperlink{https://github.com/abbew25/Measuring\_bulkflows}{https://github.com/abbew25/Measuring\_bulkflows}. The mocks of the SDSS data can be found at \hyperlink{https://zenodo.org/record/6640513}{https://zenodo.org/record/6640513}. The mock simulated data of the CF4 dataset, and its 6dFGSv/CF4TF subsets, can be shared upon request to the authors. Further evidence, plots and information for the various tests and results we present and discuss in this work can be found here: \hyperlink{https://github.com/abbew25/Measuring\_bulkflows}{https://github.com/abbew25/Measuring\_bulkflows}.



\bibliographystyle{mnras}
\bibliography{bibliography} 



\appendix

\section{Linear theory velocity covariance matrix}
\label{appendix:lineartheorycovariance}

The linear theory covariance matrix of galaxy PVs $G_{mn}$ in a survey can be calculated given information about their coordinates $\{ \theta, \phi, D(z) \}$ as follows; the matrix element corresponding to the galaxies labelled by $m$ and $n$ is given by
\begin{equation}
    G_{mn} = \frac{a^2 H(a)^2 f^2}{2 \pi^2} \int_{V_k} P_{mm}(k) W_{mn}(k) dk,
\end{equation}
where $P_{mm}(k)$ is the matter density power spectrum, $W_{mn}(k)$ is a function that depends on the angle between the galaxies $m$ and $n$ and $f$ is the growth rate of matter in the universe, $f = \frac{d \ln{D(a)}}{d a}$. $a$ is the scalefactor and $D(a)$ is the linear growth factor. $W_{mn}(k)$ can be written as \citep{ma2011peculiar}
\begin{equation}
    W_{mn}(k) = \frac{1}{3}\cos{(\alpha)}\left(j_0(kA) - 2j_2(kA)\right) + \frac{1}{A^2}j_2(kA)r_m r_n\sin^2{(\alpha)},
\end{equation}
where $j_l$ is the spherical bessel function, $\alpha$ is the angle between $\mathbf{r}_m$ and $\mathbf{r}_n$, the position vectors for the galaxies. $A$ is defined as $A = \sqrt{r_m^2 + r_n^2 - 2r_m r_n \cos{(\alpha)}}$.


\bsp	
\label{lastpage}
\end{document}